\documentclass[prx,a4paper,aps,twocolumn,superscriptaddress,10pt,bibnotes]{revtex4-1}
\usepackage{graphicx,color}% Include figure files
\usepackage{dcolumn}% Align table columns on decimal point
\usepackage{bm}% bold math
\usepackage{amsmath,amssymb,mathrsfs}
\usepackage{url}
\usepackage{hyperref}
\newcommand{\R}{\mathbb{R}}
\newcommand{\C}{\mathbb{C}}

\newcommand{\set}[1]{\mathsf{#1}}

\newcommand{\spc}[1]{\mathcal{#1}}

\def\>{\rangle}
\def\<{\langle}
\def\kk{\>\!\>}
\def\bb{\<\!\<}
\newcommand{\st}[1]{\mathbf{#1}}

\newcommand{\map}[1]{\mathcal{#1}}
\newcommand{\Tr}{\operatorname{Tr}}

\newcommand{\op}[1]{\operatorname{#1}}

\newtheorem{theo}{Theorem}

\newtheorem{lemma}{Lemma}

\def\Proof{{\bf Proof.~}}
\def\qed{$\blacksquare$ \medskip}
\newcommand{\CJ}{Choi-Jamio\l{}kowski~}

\begin{document}
    \title{
    Tsirelson bounds for quantum correlations with indefinite causal order  
    %Bounding the  quantum violation of causal inequalities
   % Tsirelson bounds for causal inequalities are violated by classical processes with indefinite input-output direction\\
    %Tsirelson bounds for quantum correlations with indefinite causal structure}
    }

    \author{Zixuan Liu}
    \affiliation{QuIC, Ecole Polytechnique de Bruxelles, C.P. 165, Universit\'e Libre de Bruxelles, 1050 Brussels, Belgium}
   % \affiliation{QICI Quantum Information and Computation Initiative, Department of Computer Science, The University of Hong Kong, Pokfulam Road, Hong Kong}
  % \affiliation{HKU-Oxford Joint Laboratory for Quantum Information and Computation}
      \author{Giulio Chiribella}
    \email{giulio@cs.hku.hk}
    \affiliation{QICI Quantum Information and Computation Initiative, School of Computing and Data Science,  The University of Hong Kong, Pokfulam Road, Hong Kong}
   \affiliation{HKU-Oxford Joint Laboratory for Quantum Information and Computation}  
    \affiliation{Department of Computer Science, University of Oxford, Wolfson Building, Parks Road, Oxford, UK}
    \affiliation{Perimeter Institute for Theoretical Physics, 31 Caroline Street North, Waterloo,  Ontario, Canada}
    
\begin{abstract}
 Quantum theory is in principle compatible with processes  that violate  causal inequalities, an analogue of Bell inequalities that constrain the correlations observed by  sets of parties operating in a definite causal  order.   Since the introduction of causal inequalities, determining their maximum quantum violation, analogue to Tsirelson's bound for Bell inequalities, has remained  an open problem.   
  Here we provide a general method for bounding the violation of arbitrary causal inequalities, establishing limits to the correlations achievable by arbitrary local experiments and by arbitrary quantum processes with indefinite causal order. 
   We prove that the maximum  violation is generally smaller than the algebraic maximum of the corresponding correlation, and determine Tsirelson-like bounds for  a class of causal inequalities including some of the most paradigmatic examples.  
   %Finally, we   provide examples of   alternative physical theories that violate  causal inequalities to their algebraic maximum. 
    Our results motivate a search for physical principles characterizing the boundary of the set of quantum  correlations with indefinite causal order. 
   %Surprisingly,  we find that the algebraic maximum of arbitrary causal inequalities can  be achieved by  a new type of  
  % processes that allow for information to flow in an indefinite temporal  direction within the parties' laboratories. 
  \end{abstract}

  \date{March 4, 2025}
  
  \maketitle

\section{Introduction}
Traditional formulations of physics  generally  assume that events take place in a well-defined causal order.  On the other hand, it has been observed that quantum theory
 is in principle compatible with scenarios where the order of events  is indefinite \cite{chiribella2009beyond,oreshkov2012quantum,chiribella2013quantum,brukner2014quantum}, a phenomenon that has potential applications to quantum information   \cite{chiribella2012perfect,araujo2014computational,guerin2016exponential,ebler2018enhanced,zhao2020quantum,felce2020quantum,gao2023measuring,zhu2023charging}   and is expected to play an integral  role in a future theory of quantum gravity \cite{hardy2007towards}.   The possibility of indefinite causal order gives rise to  an analogue of quantum nonlocality, originally shown by  Oreshkov, Costa, and Brukner (OCB)  \cite{oreshkov2012quantum}, who 
    developed a framework for studying the most general correlations arising from quantum experiments performed in a set of local  laboratories. In this scenario,  the assumption that the experiments are performed in a definite causal order  implies 
   an analogue  of Bell inequalities, called  causal inequalities.   Strikingly, OCB found that the validity of quantum theory  in the local laboratories is  compatible with correlations that violate a causal inequality, now called  the OCB inequality. 
   
 Over the past decade,   quantum violations of  causal inequalities  were found  in a variety of scenarios  \cite{oreshkov2016causal,branciard2015simplest,baumeler2014maximal,feix2016causally,abbott2016multipartite}, also including scenarios achievable with known physics using time-delocalized quantum systems \cite{wechs2023existence}.    These findings raise a fundamental question:  what is the maximum violation of causal inequalities  that is logically consistent with the validity of quantum theory in a set of local laboratories?      Can it reach the maximum algebraic value of the corresponding correlations?    Answering these questions is important for understanding the extent to which quantum mechanics is compatible with indefinite causal order. In particular, if the maximum quantum-compatible  violation  turns out to be smaller than the maximum algebraic value, then  the knowledge of the maximum quantum value provides the basis for a search for physical principles explaining why quantum theory is in principle compatible with some amount of causal indefiniteness, but not with the maximal amount. Analogues of these questions  have been extensively studied in the context of   Bell inequalities
 %, where they led to  the problem of finding  the maximum Bell violations   allowed by quantum mechanics 
\cite{cirel1980quantum,popescu1994quantum,van1999nonlocality,brassard2006limit,navascues2007bounding,brunner2009nonlocality,linden2007quantum,pawlowski2009information,navascues2010glance,fritz2013local}, where they provided valuable insights into the physical and information-theoretic principles underlying quantum theory.  
 In the paradigmatic case of the Clauser-Horne-Shimony-Holt (CHSH) inequality \cite{clauser1969proposed}, the maximum quantum violation was identified by Tsirelson \cite{cirel1980quantum} and is  now known as the Tsirelson bound.

      In stark contrast with the Bell scenario,   little is known about the boundary of the set of quantum correlations  with indefinite causal order.  %Strikingly, the maximum quantum  violation has so far remained unknown for most causal inequalities.
      %  except  those where the quantum violation is  already equal to the maximum algebraic value of the corresponding correlations.   
  For the OCB inequality,  an upper bound on the violations achieved by   a restricted class of local experiments was derived in Ref. \cite{brukner2015bounding}, but whether larger violations could be achieved by  more general experiments remained as an open question.  
  %In fact,  even the  basic question of whether or not the quantum violation of the OCB inequality  reaches its algebraic maximum has remained unanswered so far.
   For another causal inequality, associated to the Guess-Your-Neighbor's Input game \cite{almeida2010guess}, it was recently proven that the maximum violation has to be strictly smaller than the algebraic maximum \cite{kunjwal2023nonclassicality}. 
   %which raises the question of how large is the gap between the algebraic maximum and   the quantum value. For general causal inequalities, 
   In general,  however, no upper bound other than the algebraic maximum has been found for any causal inequality so far.  
%In addition, new questions have recently arisen  from  the introduction of a new class of scenarios where not only the causal order of the experiments, but also the temporal direction of the information flow within the local  laboratories can be  indefinite \cite{chiribella2022quantum}. Can these scenarios lead to even larger violations?  And in the affirmative case, where does  the boundary lie  between the correlations achievable  with indefinite causal order alone and those  achievable when indefinite causal order is combined with indefinite temporal direction? 

Here we develop   a general method for bounding the maximum violation of arbitrary causal inequalities by quantum processes with indefinite causal order.  As an application of the general method, we establish the analogue of  Tsirelson's bound for the OCB inequality and for a class of causal inequalities, which we name single-trigger inequalities. 
We then ask whether there exist alternative physical theories that allow for the maximum algebraic violation of causal inequalities. We answer the question in the affirmative, by showing two variants  of classical and quantum  theory that are  in principle compatible with correlations that reach the maximum algebraic value of all  causal inequalities with two parties, with up to three settings per party.
%We then  show that allowing information to flow in an indefinite temporal direction within the local laboratories    leads to a violation of all causal inequalities to their algebraic maximum for arbitrary numbers of parties, outcomes, and settings. Remarkably, this extreme violation of causal inequalities can be achieved  even if  all local laboratories are restricted to classical operations.  
 %Our result establishes correlations with indefinite order and time direction as the analogue of general no-signalling correlations in the Bell  scenario.  
 Overall, our findings 
    open  up a search for physical principles determining  the boundaries of the set of quantum correlations with indefinite order, potentially leading to a new axiomatization of quantum theory in that does not presuppose  a pre-defined causal structure. 
    %and a search for  potential applications  to quantum information.  

\section{Results}

{\bf Single-trigger causal inequalities.}  Here we introduce a special class of causal inequalities that provide the foundation of our method.  For the inequalities in this class, the maximum violation can be determined explicitly by a semidefinite program, which in turn  can be used to provide upper bounds to the violation of arbitrary causal inequalities.

In the framework of causal inequalities \cite{oreshkov2012quantum,oreshkov2016causal,abbott2016multipartite},    a set of parties operate in different regions of spacetime, performing local  operations  in their laboratories.  The interaction between the parties'  laboratories and the outside world takes place only at specific moments:  in the simplest presentation of the framework,  the $i$-th laboratory is assumed to be shielded from the outside world at all times, except for two moments $t_i$ and $t_i' \ge t_i$ when a shutter is opened, allowing physical systems to enter and exit the laboratory, respectively \cite{oreshkov2012quantum}. 

 In the time between $t_i$ and $t_i'$ the $i$-th party performs an experiment, obtaining an outcome.   We denote by  $x_i$ ($a_i$) the setting (outcome) of the experiment performed by the $i$-th party, and by $\vec x = (x_1, \dots, x_N)$ ($\vec a = (a_1, \dots, a_N)$) the vector of all parties'  settings (outcomes).   
An $N$-partite correlation function is an expression of the form  
\begin{align}\label{correlation}
\map I = \sum_{\vec a, \vec x} \, \alpha_{\vec a, \vec x} ~  p(\vec a \,|\, \vec x)\, ,
\end{align} 
where each $\alpha_{\vec a,\vec x}$ is a  real coefficient and $p(\vec a \,|\, \vec x)$ is the conditional probability distribution of the outcomes given the settings.

When the parties operate in a definite causal order,  the probability distribution $p( \vec a\,|\, \vec x)$ is subject to a set of linear constraints \cite{oreshkov2012quantum,oreshkov2016causal,abbott2016multipartite}.       In the  case of $N=2$ parties, Alice and Bob, the constraints have a simple expression: if Alice's experiment  precedes Bob's experiment, then the probability distribution of Alice's outcomes  must be independent of Bob's settings, namely $p_A(a_1\,|\,    x_1,x_2)   =  p_A (  a_1\,|\, x_1,x_2')$,   $\forall x_2,x_2'$, with $p_A( a_1\,|\, x_1,x_2)  :  =  \sum_{a_2}  \,  p(  a_1,a_2\,|\,   x_1,x_2)$.   Vice-versa, if Bob's experiment precedes Alice's experiment, then Bob's outcomes must be independent of Alice's settings, namely    $p_B(a_2\,|\,    x_1,x_2)   =  p_B (  a_2\,|\, x_1',x_2) \, , \forall x_1,x_1'$, with $p_B( a_2 \,|\, x_1,x_2)  :  =  \sum_{a_1}  \,  p(  a_1,a_2\,|\, x_1,x_2)$. A probability distribution $p( a_1,a_2\,|\, x_1,x_2)$ is called {\em causal} if it is a random mixture of probability distributions corresponding to  scenarios in which either Alice's experiment precedes Bob's or Bob's experiment precedes Alice's \cite{oreshkov2012quantum}.  In the multipartite case, causal probability distributions can arise in a more general way, by  dynamically controlling the order of some of the parties based on outcomes  obtained by some of the other parties  \cite{oreshkov2016causal,abbott2016multipartite}.

A causal inequality is an upper bound on the correlations achievable by causal probability distributions; explicitly, it is a bound of the form    $\map I^{\rm causal} \le \beta$, where $\map I^{\rm causal}$ is the maximum correlation achieved by causal probability distributions and $\beta\in \R$ is some constant.  The first example of a causal inequality was introduced by OCB \cite{oreshkov2012quantum}, who showed that quantum theory is in principle compatible with  its violation.

 In general, the violation of causal inequalities takes place when  the experiments performed by the different parties are connected in an indefinite order.  The connections are implemented by suitable processes, which are in principle compatible with the validity of quantum theory in the parties' local laboratories \cite{oreshkov2012quantum,chiribella2013quantum,araujo2015witnessing,chiribella2016optimal,bisio2019theoretical}.
  In the $N$-partite setting, a process  of this kind  is  represented by a linear map   $\map S$ that transforms the parties' local  operations  into the conditional probability distribution  
\begin{align}
\label{icoprob}
p  (\vec a \,|\, \vec x)   =   \map S   \left(  \map M^{(1)}_{a_1 \,|\, x_1}, \cdots, \map M^{(N)}_{a_N \,|\, x_N}    \right) \, ,
\end{align}
where, for every $i  \in  \{1,\dots, N\}$, $\map M^{(i)}_{a_i \,|\, x_i}$ is the quantum operation occurring in the $i$-th laboratory when an experiment with setting $x_i$ produces the outcome $a_i$.     The positivity and  normalization of the probability distributions $p(\vec a \,|\, \vec x)$ for every possible set of experiments performed in the parties' laboratories place constraints on the admissible  maps  $\map S$ (see Appendix \ref{app:sdp} for details.)  The set of maps $\map S$ satisfying these constrains contains all possible processes with definite causal order, as well  as  another type of processes, hereafter referred as quantum processes with indefinite causal order (ICO).  

Given a correlation function $\map I$, a fundamental problem is to determine the maximum of $\map I$ over all  possible probability distributions  (\ref{icoprob}) generated by  local quantum experiments and by   quantum processes with ICO.  In the following, we will denote the maximum by  $\map I^{\rm ICO}$, and call it the ICO bound.   
 The ICO bound  is an analogue of  the  Tsirelson bound  \cite{cirel1980quantum} for causal inequalities:   the value  of $
 \map I^{\rm ICO}$ determines whether  quantum mechanics allows for a violation of the causal inequality for the correlation $\map I$, and, in the affirmative case, it provides the  maximum  violation in principle compatible with the validity of quantum theory in the parties' local laboratories.    It is worth noting that, while the physical realization of general ICO processes is currently an open problem, an important subclass of ICO processes can be realized using time-delocalized quantum systems \cite{oreshkov2019time} and some processes in this subclass have been shown  to violate causal inequalities \cite{wechs2023existence}.  Since the ICO bound limits the violations achievable by arbitrary  ICO processes, in particular it provides an upper bound on the  maximum  violations arising from  time-delocalized quantum systems, thereby providing a non-trivial constraint on the observable consequences of  delocalization in time.

  Finding the maximum of a correlation  over all local experiments and over all ICO processes is a difficult optimization problem, which in principle requires a maximization over quantum systems of arbitrary dimensions,   as in the case of Bell correlations.  Compared to the maximization of Bell correlations, however, the maximization of ICO correlations appears to be a harder problem, and  the exact value of the ICO bound has remained  unknown until now for all causal inequalities, except those that have been shown  to be violated to their algebraic maximum \cite{baumeler2014maximal,baumeler2016space,abbott2016multipartite}.

We now introduce a class of $N$-partite correlations for which the maximum quantum violation can be computed explicitly.   These  correlations, called single-trigger correlations,  can be thought as  the score achieved by the parties  in a game where each party is asked a question and the payoff depends on the party's answer only if the party receives a specific question, called the ``trigger.''    The precise definition is as follows:  an $N$-partite correlation (\ref{correlation})
    is {\em single-trigger} if 
    for every party $i$ there exists one and only one setting $\xi_i$ (the ``trigger'')    such that   $\alpha_{\vec a,\vec x}$ depends on $a_i$ only if $x_i   =  \xi_i$.

An example of single-trigger correlation in the $N=2$ case arises from the game known  as  Lazy Guess Your Neighbor's Input  \cite{branciard2015simplest}.   This game involves  two parties,  Alice and Bob, each of which has to  guess the other party's setting when her/his own setting is equal to 1.   The probability of success is  $P_{\rm succ}^{11}  :=  p(1,1 \,|\, 1,1)$  if both parties have input  1,  $ P_{\rm succ}^{01}   :  =  p_B(  0 \,|\, 0,1)$  or  $P_{\rm succ}^{10}   :  =    p_A( 0 \,|\, 1,0)$ if only one of the parties has input 1, and  is equal to one when both parties have input 0.   Assuming uniform probabilities for the possible inputs,  the average probability of success is 
$\map I_{\rm LGYNI} =   ( P_{\rm succ}^{11}    +  P_{\rm succ}^{01} +  P_{\rm succ}^{10}  +  1)/4$.

Crucially, every correlation $\map I$ can be decomposed into a sum of single-trigger correlations, and the number of non-zero terms in the sum is  at most equal to the total number of settings, namely $ n_1  \cdot n_2  \cdots  n_N$, where $n_i$ is the number of settings for the $i$-th party.      Indeed, the existence of a decomposition with this number of terms is immediate, as one can always write the correlation   $\map I = \sum_{\vec a, \vec x} \,  \alpha_{\vec a,\vec x}\, p(\vec a \,|\, \vec x)$ as $\map I  =  \sum_{\vec \xi}   \map I_{\vec \xi}$, where $\map I_{\vec \xi}$ is the  single-trigger correlation with coefficients $\alpha^{\vec \xi}_{\vec a,\vec x}  :  =   \delta_{\vec x, \vec \xi}\,   \alpha_{\vec \xi, \vec a}$.       In general, there can be other decompositions with a smaller number of non-zero terms.   

  In the following, the causal inequalities associated to single-trigger correlations will be called ``single-trigger causal inequalities."   For these inequalities, we will provide an explicit expression of the ICO bound.      Since every possible correlation can be decomposed into a sum of single-trigger correlations, this bound   will  imply general  upper bounds on the violation of arbitrary causal inequalities.

\medskip

{\bf  Maximum quantum violation  of single-trigger causal inequalities.}      We now provide the central result of the paper: the ICO bound  for single-trigger causal inequalities. A crucial feature  of our result is that  it   reduces the problem of calculating the ICO bound to  a semidefinite program  (SDP) on a space of fixed dimension, depending only on the number of parties, settings, and outcomes.   %Thanks to this fact, one does not need to optimize over the dimension  of all possible quantum systems in  the parties' laboratories.  

In addition to the value of the ICO bound, our result provides a canonical choice of local experiments that  achieve the bound. The canonical choice  for the $i$-th party is to use as input a quantum  system of dimension $m_i$, equal to the number of possible outcomes, and to append  to it an auxiliary  system of dimension $n_i$, equal to the number of possible settings. In other words, the canonical strategy uses an input system with  Hilbert space  $\spc H^{(i)}_{\rm in}    =   \C^{m_i}$ and an output system with Hilbert space   $\spc H^{(i)}_{\rm out}    = \spc H^{(i)}_{\rm in} \otimes \spc H^{(i)}_{\rm aux}$ with $\spc H^{(i)}_{\rm aux}  =  \C^{n_i}$.    The  auxiliary system is used by the $i$-th party to communicate their setting to the outside world: explicitly, the value of the setting will be encoded in the pure state $|x_i\>\<x_i|$, where $\{  |x_i\>\}_{x_i=1}^{n_i}$ is a fixed orthonormal basis for $\spc H^{(i)}_{\rm aux}$.   If the setting is equal to the trigger,  then the party will measure the input system on the canonical basis for $\spc H_{\rm in}^{(i)}$. If the setting is not equal to the trigger, the party will not perform any measurement and will just output a random outcome.  Mathematically, this  strategy is described by quantum operations   of the form  $\map M^{(i)   *}_{a_i  \,|\,  x_i}   =  \map N^{(i)*}_{a_i\,|\, x_i}   \otimes |x_i\>\<x_i|$,  where  $\map N^{(i)*}_{a_i\,|\,   x_i}$ is the quantum operation defined by   $\map N^{(i)*}_{a_i\,|\,   x_i}  (\rho):=\<a_i|  \rho  |a_i\>  \,  |a_i\>\<a_i| $ if $x_i=  \xi_i$ and $\map N^{(i)*}_{a_i\,|\,   x_i}   (\rho):=  \rho/m_i$ if $x_i\not = \xi_i$, for every density matrix $\rho$.   

The value of the ICO bound is expressed in terms of  the canonical strategy.  To this purpose, we define the linear map
%\begin{align}\label{MI}
$\map M^*_{\map I}     =  \sum_{\vec a,\vec x}\, \alpha_{\vec a,\vec x} \,  \map M^*_{\vec a  \,|\, \vec x}$,
%\end{align}
where  $\alpha_{\vec a,\vec x}$ are the coefficients of the correlation $\map I$  [cf. Eq. (\ref{correlation})]  and $\map M^*_{\vec a  \,|\, \vec x}  : = \map M^{(1)*}_{a_1  \,|\,  x_1} \otimes \map M^{(2)*}_{a_2  \,|\,  x_2} \otimes \cdots \otimes \map M^{(N)*}_{a_N  \,|\,  x_N}$ are the parties' local operations.   
We then prove that the ICO bound is  a measure of  the deviation between the map $\map M^*_{\map I}$ and the set of $N$-partite no-signalling channels, that is, the set of quantum processes that do not allow any subset of the  parties  to transmit information to any other subset  (see Methods).    Without loss of generality,  here we   focus our attention on the case where all the coefficients $\alpha_{\vec a,\vec x}$ are nonnegative. 
   In this case,  we show that the single-trigger ICO bound is 
    \begin{align}\label{triggertsirelson}
  \map I^{\rm ICO}_{\rm single-trigger}     =    2^{D_{\max}(\map M^*_{\map I} \, \| \, \set{NoSig})} \, , 
   \end{align} 
 where   $\set{NoSig}$ is the set of $N$-partite no-signalling channels,  $D_{\max}  (\map A \|  \map B):= \min \{  \lambda~|~  \map  A  \le   2^\lambda\,  \map B     \}$ is the max relative entropy between two completely positive maps $\map A$ and $\map B$ (with the notation $\map A \le \map B$ meaning that $\map B-\map A$ is completely positive),  and, for a set of completely positive maps $\set S$,  $D_{\max}  ( \map A\|     \set S   )  :=  \max_{\map B  \in  \set S}  D_{\max}  (\map A\|  \map B)$.     
 
 The quantity  $D_{\max}(\map M^*_{\map I} \, \| \, \set{NoSig})$ appearing in the ICO bound is also known as the max relative entropy of signalling \cite{chiribella2016optimal}.  
 Eq. (\ref{triggertsirelson})  provides an SDP expression for the ICO bound  of all single-trigger causal inequalities.     In particular, it  yields the ICO bound  $\map I_{\rm LGYNI}^{\rm ICO}  \approx  0.8194$  for the single-trigger causal inequality corresponding to the LGYNI game.  This result provides the answer to  an open question  raised in Ref. \cite{branciard2015simplest}, where the value $\map I_{\rm LGYNI}   \approx  0.8194$   was obtained  by optimizing over  local experiments  and ICO processes  under the assumption that all quantum systems in the parties' local laboratories are  two-dimensional.

\medskip 

{\bf Bound for arbitrary causal inequalities}.     The ICO bound for single-trigger causal inequalities directly implies a general bound for arbitrary causal inequalities. The bound follows from the fact that every correlation $\map I$ can be decomposed as a sum of single-trigger correlations, and therefore its ICO bound cannot exceed the sum of the ICO bounds for the single-trigger correlations in its decomposition.  
   Optimizing over all possible decompositions, one obtains the bound 
  \begin{align}\label{generalbound}
 \map I^{\rm ICO}   \le   \min_{ \left\{ \map I_{j} \right\}_{j=1}^n     }   \left\{  \sum_{j=1}^n   \,   \map I_{j}^{\rm ICO}\right\}\, ,
 \end{align} 
where   $\left\{\map I_{j}\right\}_{j=1}^n$ are  single-trigger correlations satisfying the condition $\sum_{j=1}^n  \,  \map I_j  =  \map I$.

In Appendix \ref{app:sdp}, we show that  the evaluation of the right-hand-side of Eq. (\ref{generalbound})  is an SDP on a space of fixed dimension, and provide a characterization of this SDP  in terms of a set of necessary conditions on quantum correlations with indefinite causal order.   The solution of the SDP  provides a computable upper bound on the violation of causal inequalities for every desired correlation $\map I$.

In the next sections, we show three important implications of our  general results.  

\medskip  
{\bf Maximum quantum violation of the OCB inequality.}  The first  example of a causal inequality  is the OCB inequality \cite{oreshkov2012quantum}, which corresponds to a  bipartite scenario  in which $a_1,a_2,$ and $x_1$ are bits, while $x_2$ is a pair of bits, here denoted by $b$ and $c$, respectively. The inequality arises from  a game, in which either Alice is asked to guess Bob's bit  $b$  or Bob is asked to guess Alice's bit $x_1$,  depending on the value of bit $c$.   When the order between Alice's  and Bob's  experiments is well-defined,  only one player can communicate the value of his/her input to the other player, and therefore one of the two players must make  a random guess of the other players' input.  Hence, every causal probability distribution satisfies the constraint that the probability that Alice's guess is correct, plus the probability that Bob's guess is correct, is upped bounded as
%\begin{equation}\label{OCB}
 ${\map I}_{\rm  OCB}^{\rm causal}    :  =  P(a_1=b \,|\, c=0) + P (a_2=x_1 \,|\, c=1)   \le \frac 32$, 
%\end{equation}
with $P(a_1=b \,|\,  c=0)    :=\sum_{x_1,b}  \,  p_A(b \,|\,  x_1,b,0)/4 $ and $P(a_2=x_1 \,|\,  c=1)    := \sum_{x_1,b}  \,  p_B(x_1\,|\, x_1, b, 1)/4$. 

OCB showed that the  above causal inequality  can be violated by a quantum ICO process and by suitable experiments in Alice's and Bob's laboratories, the combination of which reaches the value 
\begin{align}\label{OCBmax}
{\map I}_{\rm  OCB}   = 1  +  \frac 1  {\sqrt 2}  \,.
\end{align}
  A fundamental question raised in the original OCB paper  is whether higher violations are possible, and, in case they are, what is the maximum violation.  %Finding the maximum violation of the OCB inequality  is the analogue of deriving the  Tsirelson bound for the CHSH inequality in the Bell scenario \cite{cirel1980quantum}. 
 The value (\ref{OCBmax}) was shown to be maximum among the violations achievable with  a restricted set of local measurements, involving measurement and repreparations with traceless binary observables   \cite{brukner2015bounding}. Whether more general types of measurements  could lead to higher violations, however,  remained as an open question until now.  In fact,  even the most basic question of whether the violation of the OCB inequality over arbitrary measurements and arbitrary ICO processes  can reach its maximum algebraic value $\map I_{\rm OCB}  = 2$  had remained unanswered until the present work. 
 
We  now  solve the problem in full generality, proving that (\ref{OCBmax}) is indeed the largest violation allowed by arbitrary quantum processes with indefinite causal order and by arbitrary operations in Alice's and Bob's laboratory. 
In fact, we derive the exact ICO bound for a version of the OCB correlation, called the biased OCB correlation  \cite{bhattacharya2015biased} and given by  
\begin{align}
\map I_{{\rm OCB}, \alpha} = P(a_1=b \,|\, c=0) + \alpha \, P(a_2=x_1 \,|\, c=1)\, ,
\end{align}
where $\alpha$ is an arbitrary real number.  
  To derive the ICO bound,  we observe that the biased OCB correlation is a random mixture of two single-trigger correlations    and we evaluate the ICO bound (\ref{triggertsirelson})  for these two correlations.   In this way, we obtain an upper bound on the quantum violation,  which we show to  coincide with the value  found in  \cite{bhattacharya2015biased} with   a specific example of  ICO process   (see Appendix \ref{app:ocblemma}). All together,  these results establish the ICO bound 
   \begin{align}\label{OCBalpha}
      \map I_{\rm OCB,  \alpha}^{\rm ICO} = \frac{1 + \alpha + \sqrt{1+\alpha^2}}{2} \, ,
        \end{align} 
thereby providing an analogue of the  Tsirelson bound for indefinite causal order.

\begin{figure}[htbp]
  \centering
    \includegraphics{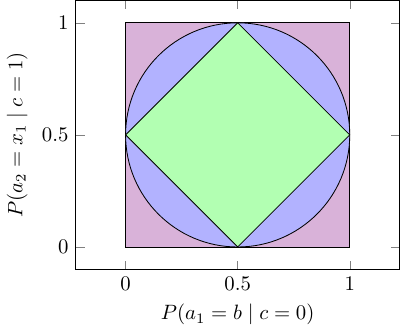}
    \caption{{\bf Geometry of causal, ICO, and general probability distributions.}  The figure shows the two probabilities appearing in the OCB correlation.  The inner square (in green)  corresponds to the values achievable by causal probability distributions, while the  circle (in blue) corresponds to the values achievable through quantum ICO processes.  The outer square  (in violet) corresponds to   the values achievable by arbitrary, unconstrained probability distributions.}
    \label{fig:geometry}
\end{figure}

 \medskip
{\bf The quantum ICO set of correlations.}     Our results provide  insights into the geometry of the set of probability distributions generated by quantum ICO processes.     Let us consider the paradigmatic case of the OCB  game, and visualize the possible values of Alice's and Bob's success probabilities  $P_A  : =  P(a_1=b \,|\, c=0)$ and $P_B:  =P(a_2=x_1 \,|\, c=1)$ by representing them in a two-dimensional plane.       With this notation, the  ICO bound Eq. (\ref{OCBalpha}) is equivalent to the condition  $  \cos \theta \, (P_A-1/2)    + \sin \theta \,    (P_B-1/2)  \le  1/2$ for $\theta  =   \arctan \alpha$.            This condition identifies  a circle of radius $1/2$ centred around the point $(1/2,1/2)$, as illustrated in Figure \ref{fig:geometry}.     In the figure, we also show the area corresponding to causal probability distributions,  
 which form a square inscribed inside the circle.   Finally, arbitrary unconstrained probability distributions  occupy the square circumscribing the circle of quantum ICO probabilities. 
 
  Intriguingly, Figure \ref{fig:geometry}  coincides with the  analogous picture in the case of the CHSH inequality, where the smaller square is the set of probability distributions allowed by local realism, the circle is the set of probability distributions allowed by quantum mechanics, and the larger square is the set of general no-signalling probability distributions \cite{brunner2014bell}.   Later in this paper, we will show that, in the case of causal inequalities, the largest square corresponds to probability distributions achievable in an alternative physical theory where the local parties are restricted to a subset of operations and the processes connecting the parties' experiments are only required to satisfy the positivity and normalization of probabilities for operations in  this restricted subset.
  
  %   generated by local experiments 
 %time-symmetric variants of  classical theory.  
  %a new class of  operations that do not assume a definite direction of time outside Alice's and Bob's laboratories. 

In Appendix \ref{app:tiltedlgyni},  we provide further insights into the geometry of quantum correlations with indefinite causal order by analyzing the set of  correlations arising from the LGYNI game and a biased version thereof.

\medskip  

{\bf Bound for   Guess Your Neighbor's Input.}    Guess Your Neighbor's input  (GYNI) \cite{almeida2010guess} is a multipartite quantum game where each party has to guess the input of one of its neighbors.      This game is well known in the study of quantum nonlocality, being the first example of a tight Bell inequality with no quantum violation.      In the two-party setting, 
 the probability of success is
%\begin{equation}
    $\map I_{\rm GYNI} = P(a_1=x_2, \, a_2=x_1)$. 
%\end{equation}
GYNI  has also been studied  in the context of causal inequalities \cite{branciard2015simplest,bavaresco2019semi}, where it was observed that causal probability distributions satisfy the bound $\map I^{\rm causal}_{\rm GYNI}   \le 1/2$ when the settings are uniformly random, while quantum ICO processes can violate this bound.    The question about the maximum violation of the GYNI causal inequality has been tackled in Refs. \cite{bavaresco2019semi,kunjwal2023nonclassicality}. In particular, Ref. \cite{kunjwal2023nonclassicality}  proved that the quantum ICO violation cannot reach the algebraic maximum     $\map I_{\rm GYNI} = 1$.   However, no explicit bound on the maximum violation other than the algebraic maximum has been known up to now.    

Using our general expression (\ref{generalbound}),  we can now show that the ICO bound for  the GYNI game satisfies the inequality  $\map I_{\rm GYNI}^{\rm ICO}  \le 0.7592$. At present, this inequality represents the state-of-the-art in upper bounding the quantum ICO violation of the  GYNI inequality.   Nevertheless, finding the  exact value of $\map I_{\rm GYNI}^{\rm ICO}$ remains  an open problem. Previous numerical results based on  a see-saw algorithm using quantum systems of dimension 5 showed that quantum ICO processes can reach the value   $\map I_{\rm GYNI} \approx   0.6218$ \cite{branciard2015simplest}.  An interesting question is whether increasing the dimension of the local quantum systems could yield correlations that go all the way to the upper bound   $0.7592$, or whether the maximum quantum violation is attained at a strictly smaller value.

\medskip

{\bf Reaching the algebraic maximum.}  We have seen that the maximum violation of causal inequalities compatible with standard quantum theory is subject to nontrivial constraints. A natural question is whether  it is possible to find   alternative physical theories in which the violation can in principle reach its algebraic maximum. In the context of Bell inequalities, this question has been asked   by Popescu and Rorlich \cite{popescu1994quantum}, who showed that the algebraic maximum of the CHSH inequality can be achieved in an alternative theory, now known  as boxworld \cite{gross2010all}.   For causal inequalities,  we  provide two examples of  theories that reach the algebraic maximum of the GNYI inequality and other two-party causal inequalities. 
%,   based 

Consider a  variant of quantum theory  in which the experimenters  are restricted to perform only bistochastic instruments, that is, instruments $(\map M_{a \,|\, x})_{a=1}^m$ satisfying the condition that the map $\sum_a  \map M_{a \,|\, x}$ is both trace-preserving and identity preserving.      We call this theory bistochastic quantum theory 
 \cite{chiribella2021symmetries}.   As in ordinary quantum theory,  one can define the set of all logically conceivable supermaps in bistochastic quantum theory.  
  These supermaps,  introduced in Ref. \cite{chiribella2022quantum},  
 %where they were called quantum operations with indefinite time direction. 
  represent the most general processes that could in principle connect the local operations performed by parties obeying to bistochastic quantum theory,  giving rise to probability distributions as in Eq.   (\ref{icoprob}), where now all the instruments are required to be bistochastic.  In the two-party scenario,    an explicit  characterization of the admissible supermaps is provided  in Methods.

The above construction also applies to the bistochastic  version of classical theory,    which  can be obtained from bistochastic quantum theory by  subjecting every system to a decoherence map.    Explicitly, a bistochastic classical channel is described by a conditional probability distribution specifying  the probability $q^{(i)}  (s_i'\,|\, s_i) \ge 0$ that an input state $s_i$ at time $t_i$ is transformed into an output state $s_i'$ at time $t_i'$, and satisfying 
%Compatibility with two alternative directions  of the information flow requires the probability distribution to be bistochastic, that is, to 
the  conditions $\sum_{s_i'}  q^{(i)}  (s_i'\,|\,  s_i)  =  1$ and  $\sum_{s_i}  q^{(i)}  (s_i'\,|\,  s_i)  =  1$.      In classical bistochastic theory, the most general experiments performed by the $i$-th party  correspond to classical bistochastic instruments, mathematically described by subnormalized probability distributions $   q^{(i)}_{a_i  \,|\,  x_i}  (s_i' \,|\, s_i)$ satisfying the condition that $  \sum_{a_i} q^{(i)}_{a_i  \,|\,  x_i} (s_i' \,|\, s_i)$ is  a classical bistochastic channel  for every setting $x_i$.   The set of logically conceivable supermaps in this theory coincides with the decohered version of the supermaps allowed in bistochastic quantum theory   (see  Methods and Appendix \ref{app:classify} for more details).  
% The most general processes connecting the parties' experiments are described by admissible classical supermaps  (see  Methods for more details).  

In the following, the  maximum of a correlation $\cal I$ over the dimensions of the parties' systems, over all choices of bistochastic instruments, and over all admissible supermaps will be called the bistochastic ICO (BICO)  value.  The main result of this section is that the quantum and classical BICO values coincide and are equal to the algebraic maximum  for every  two-party correlation with up to three settings.  To prove this result, we start from the GYNI game, which can be won with certainty only if Alice and Bob can perfectly signal to one another.    

Consider a scenario where all the  systems  entering and exiting the parties' laboratories are classical bits. To win the GYNI game, Alice and Bob adopt the following strategy:  First, they measure the values of the bits $s_1$ and $s_2$ entering  their laboratories, respectively, and, depending on their settings, they either leave the bit unchanged or they filp them, obtaining the new values $s_1'    =  s_1 \oplus x_1 $ and $s_2'  =  s_2\oplus x_2$, respectively (here $\oplus$ denotes addition modulo 2).   The final value of the bit then becomes the outcome of their measurements, namely $a_1  =   s_1'$ and $a_2  = s_2'$.     Mathematically, this strategy  corresponds to the classical bistochastic instruments  $q^{(1)}_{a_1 \,|\, x_1}    (s_1'  \,|\,  s_1)   =  \delta_{s_1'  ,  s_1  \oplus x_1}  \,  \delta_{a_1,  s_1'} $ for Alice and   $q^{(2)}_{a_2 \,|\, x_2}    (s_2'  \,|\,  s_2)   =  \delta_{s_2'  ,  s_2  \oplus x_2}  \,  \delta_{a_2,  s_2'}$ for Bob.    To achieve unit probability of winning  the GYNI game,  Alice's and Bob's local  experiments  are connected by a  deterministic process  $\map S$,   represented as a function that  maps the final values of the bits $(s_1',s_2')$ into their initial values $(s_1,  s_2)$ according to the rule 
\begin{equation}
    \label{eq:processfunc}
    s_1= s_1' \oplus s_2'  \qquad {\rm and} \qquad    s_2= s_1' \oplus s_2' \, .
\end{equation}
The action of the process $\map S$  on the parties' operations produces the probability distribution  $\map S   ( q^{(1)}_{a_1\,|\, x_1},    q^{(2)}_{a_2\,|\, x_2} )$  $=\sum_{s_1', s_2'}    \,   q^{(1)}_{a_1\,|\, x_1}   (s_1' \,|\, s_1'\oplus s_2')  \, q^{(2)}_{a_2\,|\, x_2}   (s_2'\,|\, s_1'\oplus s_2') 
%~\delta_{s_1, s_1'\oplus s_2'} \, \delta_{s_2,  s_1'\oplus s_2'}
$.    In Methods, we show that the process $\map S$ is admissible, in the sense that it gives rise to a normalized probability distribution for every choice of bistochastic instruments $q^{(1)}_{a_1\,|\, x_1}$ and $ q^{(2)}_{a_2\,|\, x_2}$.  
% Moreover, we show that the process $\map S$, regarded as a quantum process, is also a valid supermap in bistochastic quantum theory.    
 For the specific choice of instruments described earlier in this paragraph,   the process $\map S$ generates the perfect signalling distribution   $p(a_1,a_2\,|\, x_1,x_2) =  \delta_{a_1, x_2} \,  \delta_{a_2,x_1}$, which allows Alice and Bob to win the GYNI game with certainty.  

%Notice that the perfect signalling  achieved in classical bistochastic theory can used to win not only the GYNI game, but also to violate every bipartite causal inequality up to its algebraic maximum.   %This finding shows a striking difference between indefinite causal order alone and indefinite causal order combined with indefinite time direction:  in the ICO case it is known that no classical bipartite process can violate any causal inequality \cite{oreshkov2012quantum}, whereas in the ICOTD case there exists a bipartite process can reach  the algebraic maximum violation for all causal inequalities.    
 In Appendix \ref{app:omnipotent},  we extend the above result from GYNI to arbitrary causal inequalities  with two parties and up to three settings.  Specifically, we prove that perfect two-way signalling of ternary digits is achievable by choices of classical bistochastic instruments and  classical  supermaps. Equipped with a perfect two-way signalling of each other's setting, the two parties can generate arbitrary deterministic conditional probability distributions, and therefore can reach the algebraic maximum for every given correlation function.   In addition, we show that if each party has only two settings,  then every bipartite conditional probability distribution can be generated by a suitable choices of  instruments and    supermaps in classical bistochastic theory.
% As a corollary, we obtain that  the classical BICO value is equal to the algebraic maximum for all two-party causal inequalities with up to three settings per party. 
  Since  instruments and  processes in bistochastic classical theory are special cases of instruments and processes in bistochastic quantum theory, the same results hold in the quantum case.

\section{Discussion}

%Our results offer new insights into the structure the set of quantum correlations generated by quantum ICO processes. 
% and can be used as a tool to better understand the operational implication of ICO in quantum theory.  
  Our general bound on the violation of  causal inequalities is based on  an SDP relaxation of  the original problem of computing the quantum ICO bound.    For  certain correlations, such as the OCB correlation and all single-trigger correlations, we have shown that the SDP gives the exact value of the quantum ICO   bound.  In principle, one could ask whether our bound is tight for all possible   causal inequalities.  The analogy with Bell inequalities, however,  suggests a negative answer.   In Bell scenarios, a converging sequence of upper bounds on the value of maximal quantum violations  is provided  by the Navascu\'es-Pironio-Ac\'in SDP hierarchy   \cite{navascues2007bounding,navascues2008convergent}.    The analogy with this situation  suggests that our SDP relaxation may be just the first level of a a similar hierarchy of SDPs.  Determining whether this analogy is correct,  and, in the affirmative case, identifying the other levels of the hierarchy are among the most important research directions opened by our work.  Another interesting direction is to extend our method for the calculation of the ICO bound to other type of inequalities with non-trivial causal structure, such as the inequalities recently studied in Refs. \cite{gogioso2023geometry,van2023device}.  
  
  A natural development of our research  is to establish self-testing results for causal inequalities, in analogy to self-testing for   Bell inequalities \cite{mayers1998quantum,mayers2004self}. For example, it is interesting to determine whether the OCB process is the only quantum process (up to local transformations) that achieves the maximum violation of the OCB inequality.   
  %Such a self-testing result may have  cryptographic implications, in a similar way as it was observed in the setting of Bell correlations.   
  More broadly, it is interesting to search for applications of the violation of causal inequalities to quantum technologies, in analogy to the  applications of the violation of Bell inequalities in  
 in quantum cryptography \cite{acin2006bell,colbeck2009quantum,pironio2010random}, communication complexity \cite{buhrman2010nonlocality},  and control of quantum hardware \cite{reichardt2013classical}.   While the physical realization of  general  quantum ICO  processes  is still an open problem, exploring the potential applications of causal inequalities  contributes to a deeper  information-theoretic  understanding of indefinite causal order in quantum theory.  

Finally, our results open up a search for physical principles capable of explaining why the violation of causal inequalities by  quantum ICO  processes is not equal, in general, to the algebraic maximum, and, of determining the exact value of the quantum violation.  In the context of Bell inequalities, the analogue question was originally raised by Popescu and Rohrlich \cite{popescu1994quantum}, and led to the discovery of new information theoretic principles, such as non-trivial communication complexity \cite{van1999nonlocality,brassard2006limit,brunner2009nonlocality},  non-trivial nonlocal computation \cite{linden2007quantum},  information causality \cite{pawlowski2009information}, macroscopic locality \cite{navascues2010glance}, and local orthogonality \cite{fritz2013local}.   The formulation of new principles that capture the quantum violation of causal inequalities is likely to be valuable for the understanding of quantum physics in spacetime, and may eventually lead to a new formulation of quantum theory that does not require a pre-defined causal structure. In turn, such a formulation may  offer guidance in the long-standing problem of unifying quantum theory with Einstein's theory of relativity.

\section{Methods}

{\bf Labelled projective instruments.}  We now develop a way   to reduce the search over the set of  all possible local experiments to a  search over a smaller set, generated by ideal measurements and ideal state preparations.   This reduction plays a similar role as the reduction to projective measurements in  Bell scenarios.

In the standard framework of quantum theory,  an experiment with $m$ possible outcomes is described by a quantum instrument \cite{davies1970operational,ozawa1984quantum,heinosaari2011mathematical}, that is, a collection of completely positive maps $ (\map M_a)_{a=0}^{m-1}$ satisfying the condition that their sum is trace-preserving.    Each map $\map M_a$ describes a physical process transforming an input system with Hilbert space $\spc H_{\rm in}$ into a (possibly different) output system, with Hilbert space $\spc H_{\rm out}$.     In the special case $\spc H_{\rm in}  =  \spc H_{\rm out}$, an instrument $  (\map M_a)_{a=0}^{n-1}$ is called projective if  each map $\map M_a$ is of the form $\map M_a (\rho)   = P_a \rho  P_a$, where $P_a$ is a projector and the projectors $(P_a)_{a=0}^{m-1}$ form a resolution of the identity.

We now introduce the notion of labelled projective instruments, that is, instruments that output a label describing their settings.  For every setting $x$, let $ (\map N_{a \,|\, x})_{a  =0}^{m-1}$ be a projective instrument on a quantum system with Hilbert space  $\spc H_{\rm in}$, and let $(\rho_x)_{x  = 0}^{n-1}$ be a set of perfectly distinguishable density matrices for an auxiliary quantum system with Hilbert space $\spc H_{\rm aux}$.  A labelled projective instrument is an instrument $ (\map M_{a \,|\, x})_{a  =0}^{m-1}$   of the form  $\map M_{a\,|\, x}  =  \map N_{a\,|\, x} \otimes \rho_x$.    

Our first result is that  the optimization of a conditional probability distribution $p(\vec  a \,|\, \vec x )$ over arbitrary  instruments and arbitrary ICO processes can be restricted without loss of generality to an optimization over labelled projective instruments.  The key result is the following theorem:

\begin{theo}
\label{theo:labeled_projective}
For every set of   quantum instruments $\big(\map M^{(i)}_{a_i|x_i}\big)_{a_i  =0}^{m_i-1}$, $x_i\in \{0,\dots,  n_i-1\}$, $i\in  \{1,\dots,  N\}$ and every process $\map S$ acting on them,  there exists a set of labelled projective instruments   $\big(\map M^{(i)\prime}_{a_i|x_i}\big)_{a_i  =0}^{m_i-1}$,   $x_i\in \{0,\dots,  n_i-1\}$, $i\in  \{1,\dots,  N\}$  (with possibly different input and output systems)    and a process $\map S'$ acting on them such that 
\begin{align} 
\map S  (  \map M^{(1)}_{a_1|x_1},  \dots ,  \map M^{(N)}_{a_N|x_N})   =   \map S'  (  \map M^{(1)\prime}_{a_1|x_1},  \dots ,  \map M^{(N)\prime }_{a_N|x_N})     
\end{align} 
for all settings $\vec x$ and all outcomes $\vec a$. 
In addition, the labelled projective measurements associated to each party $i$ can be chosen without loss of generality to have  projectors of the same rank $r_i$ for every outcome $a_i$ and every setting $x_i$.  
\end{theo}
The proof uses Ozawa's  dilation theorem \cite{ozawa1984quantum} for quantum instruments, combined with a use of  auxiliary systems to include unitary state changes depending on the settings. The details of the proof are provided  in Appendix \ref{app:labelledproj}.

Theorem \ref{theo:labeled_projective} can be used in every problem involving the joint optimization of local instruments and global ICO processes.  In particular, it can be used to simplify the search for the ICO bound of arbitrary causal inequalities.  %Later in this Methods section, 
 In the next section, we will show that a  strengthening of Theorem \ref{theo:labeled_projective} can be provided in the case of single-trigger causal inequalities.
  %Before providing  this strengthening, however, it is convenient to analyze more in detail the structure of the optimization problem involved in the calculation of the ICO bound.  

\medskip  
{\bf The canonical instrument.}  We now show that the  ICO bound for  single-trigger correlations  can be achieved  by the canonical choice of labelled projective instruments $\map M_{a_i \,|\, x_i} ^{(i)*}$ provided in the main text.   

%To derive  this result, we need to inspect more closely the structure of optimization problem. 
In general, the calculation of the ICO bound involves a double optimization, over the all possible local instruments and over all possible ICO processes.    As an intermediate step, it is useful to consider the simpler problem where the local instruments are fixed and the optimization runs over the set of  ICO processes.  Using Eqs. (\ref{correlation}) and (\ref{icoprob}), the  correlation achieved  by an ICO process $\map S$ can be written as  $\map I  =   \map S  (\map M_{\map I})$, with   $\map M_{\map I}   = \sum_{\vec a, \vec x} \alpha_{\vec a,\vec x}\, \map M_{a_1 \,|\, x_1}^{(1)} \otimes \cdots \otimes \map M_{a_N \,|\, x_N}^{(N)} $.   The maximum correlation  achieved by arbitrary processes is given by $\upsilon  (\map M_{\map I}) :  =  \max_{\map S}  \map S  (\map M_{\map I})$.  The maximization is a semidefinite program,  whose solution can be equivalently computed  as   \cite{chiribella2016optimal}   
\begin{align}\label{dualsdp}
\upsilon  (\map M_{\map I})   =  \min  \Big\{  \eta \in \R ~|~  \exists  \map C \in  \op{Aff}(\set{NoSig})   :\,         \eta \,  \map C  \ge  \map M_{\map I}  \Big\} \, ,
\end{align}
where  $\op{Aff}(\set{NoSig})$ is the set of affine combinations of  no-signalling channels, and, for two linear maps $\map A$ and $\map B$ with the same input and output spaces, $\map A  \ge \map B$  means that $\map A- \map B$ is completely positive.  
% $\big(\map M^{(i)}_{a_i|x_i}\big)_{a_i  =0}^{m_i-1}$, $x_i\in \{0,\dots,  n_i-1\}$, $i\in  \{1,\dots,  N\}$ 
   When the coefficients $\alpha_{ \vec a,\vec x}$ are nonnegative,  the minimization can be restricted without loss of generality to the set of no-signalling channels,  and the minimum is given $2^{D_{\max}(\map M_{\map I } \| \set{NoSig})}$, where  $D_{\max}(\map M_{\map I } \| \set{NoSig})$ is the max relative entropy distance between  the operator $M_{\map I}$ and the set of no-signalling channels.   In short, the maximum correlation achievable with a fixed set of local instruments is given by the deviation  of the map $\map M_{\map I}$ from the set of no-signalling channels.

We are now ready to tackle the full problem of computing the ICO value.    With the above notation, the ICO value can be written  as $\map I^{\rm ICO}  =  \max_{\map M_{\map I}}  \upsilon (\map M_{\map I})$, where the maximization is over all maps $\map M_{\map I}$ generated by all possible local quantum instruments with inputs and outputs of arbitrary dimensions.  By Theorem \ref{theo:labeled_projective}, the maximization can be restricted without loss of generality to maps $\map M_{\map I}$ generated by labelled projective instruments.  
%  A key observation is that, for fixed input/output dimensions, $\upsilon$ is a convex function of the map $\map M_{\map I}$.    
   In Appendix \ref{app:performanceop}, we show that, for single-trigger correlations and labelled projective instruments, the map $\map M_{\map I}$ can be decomposed into a convex combination of maps associated to instruments of an even simpler form, as stated by the following theorem:

\begin{theo}
    \label{lem:performanceop}
Every map $\map M_{\map I}$  associated to a given single trigger correlation and  a given set of labelled projective instruments can be  decomposed into a convex combination $  \map M_{\map I}  = \sum_j  \,  p_j  \, \map M_{\map I,  j}$, where $(p_j)$ is a probability distribution and, for every $j$,   $\map M_{\map I,  j}$ is the map associated to local  instruments 
%  $\big(  \map M_{a_i|  x_i}^{(i)}\big)_{a_i  =  0}^{m_i-1}$ 
of the form  
    \begin{equation}\label{laststep}
    \map    M_{\,a_i|  x_i}^{(i  , j)}  (\rho) = \begin{cases}
               P_{a_i,\xi_i}^{(i)} \, \rho  \,   P_{a_i,\xi_i}^{(i)} \otimes |\xi_i\>\< \xi_i| & x_i = \xi_i \, , \\
            \frac 1 {m_i}  \,  U_{x_i}^{(i,j)}  \,\rho \,  U_{x_i}^{(i,j)\dag}    \otimes |x_i\>\< x_i| & x_i \neq \xi_i \, .
        \end{cases}
    \end{equation}
    where $   \big(P_{a_i  |x_i}^{(i)} \big)_{a_i=0}^{m_i-1}$ are the projectors appearing in the original labelled projective instrument,  and  $U_{x_i}^{(i,j)}$  is a unitary operator for every $i,j,$ and $x_i$.
\end{theo}
Operationally, the instruments appearing in Theorem  \ref{lem:performanceop} describe experiments where the $i$-th player performs a labelled projective instrument when their setting is the trigger, and, for all the other settings,  the player generates a uniformly random outcome, while performing a unitary gate on the  system entering their laboratory.   Since $\upsilon$ is a convex function  of the map $\map M_{\map I}$, Theorem  \ref{lem:performanceop} guarantees that its maximization can be restricted without loss of generality to maps generated by instruments of this form.  Note that the canonical instruments $( \map M^{(i) *})_{a_i|x_i}$ are a special case of instrument of the form (\ref{laststep}), corresponding to the case where all projectors are rank-one, and all unitary operators are equal to the identity operator.

In Appendix \ref{app:isomorphism} we show that every value of the function $\upsilon  (  \map M_{\map I})$ achievable by instruments of the form  (\ref{laststep}) can be achieved by the canonical instrument.  
 Intuitively, the argument is as follows:  first, Theorem \ref{theo:labeled_projective}  guarantees that all the instruments associated to the same party  have projectors of the same rank. For single-trigger correlations,  Theorem  \ref{lem:performanceop} ensures that one can effectively consider only one projective instrument per party.  But then, a single projective instrument with projectors of equal  rank can be reduced to a rank-one projective instrument by appending an additional quantum system in the party's local laboratory, without performing any operation on it. Without loss of generality, the additional system can be incorporated in the definition of the process $\map S$ connecting the local laboratories. Similarly, arbitrary unitary operations depending on the parties' settings can be obtained from the identity operation, by appending to the process $\map S$ a controlled unitary operation, controlled by the state of the auxiliary systems that carry the values of the parties' settings (see Appendix \ref{app:isomorphism} for the details).  The above argument implies the  bound $\upsilon  (\map M_{\map I}^*)  \ge  \upsilon  (  \map M_{\map I})$ for every map $\map M_{\map I}$ generated by local instruments, and therefore $\map I^{\rm ICO}_{\rm single-trigger}  = \upsilon  (\map M_{\map I}^*).  $

  Using Eq.  (\ref{dualsdp}) we then obtain an explicit SDP expression for the ICO value.  When the coefficients $\alpha_{\vec a , \vec x}$ are nonnegative, this expression reduces to the max relative entropy distance, in agreement with Eq. (\ref{triggertsirelson}) in the main text.

{\bf Admissible supermaps in bistochastic quantum and classical theory.}    Here we consider  time-symmetric versions of quantum and classical theories, in which local experiments are described by bistochastic instruments, and we characterize the set of all logically conceivable supermaps.    
  In the quantum case, this set of supermaps was introduced in Ref. \cite{chiribella2022quantum} and were shown to be in one-to-one correspondence with a subset of positive semidefinite operators.  In the two-party case,  the logically conceivable supermaps admit a simple characterization in terms of their Choi operators: denoting by $A_{\rm in}$  and $A_{\rm out}$  ($B_{\rm in}$  and $B_{\rm out}$) the input and output systems of  the first (second) party,  an admissible supermap is described by an operator $S \in  L (  \spc H_{A_{\rm in}} \otimes \spc H_{B_{\rm out}}  \otimes \spc H_{B_{\rm in}}  \otimes \spc H_{B_{\rm out}})$ satisfying the conditions (see Appendix \ref{app:classify} for a proof)
  \begin{align}
    &S \geq 0 \, , \nonumber \\
    &\Tr(S) = d_Ad_B \, , \nonumber \\
    &{}_{[(1-A_{\rm in})(1-A_{\rm out})B_{\rm in}B_{\rm out}]}S = 0 \, ,  \label{constraints} \\
    &{}_{[(1-B_{\rm in})(1-B_{\rm out})A_{\rm in}A_{\rm out}]}S = 0 \, , \nonumber \\
    &{}_{[(1-A_{\rm in})(1-A_{\rm out})(1-B_{\rm in})(1-B_{\rm out})]}S = 0 \, , \nonumber
\end{align}
   where $\spc H_X$ denotes the Hilbert space of system $X$, and $L(\spc H)$ represents the set of linear operators on the Hilbert space $\spc H$. We use the notation ${}_{[X]}S := \Tr_X[S] \otimes \frac{I_X}{d_X}$ for a system $X$ of dimension $d_X$, and the shorthand ${}_{[\sum \alpha_X X]}S := \sum_X \alpha_X \cdot {}_{[X]}S$ for a linear combination of the terms $\{ {}_{[X]}S \}_X$ associated with a collection of systems indexed by $X$. For instance, ${}_{[(1-X)Y]}S$ serves as a shorthand for ${}_{[Y]}S - {}_{[XY]}S$.
   When the parties perform local instruments, respectively, the probability in Eq. (\ref{icoprob}) is given by  $p(  a_1,a_2|x_1,x_2)   =   \Tr \left[  S  \,  (  M^{(1)}_{a_1|x_1}  \otimes  M^{(2)}_{a_2|x_2}  )\right]$.   
  
 Admissible supermaps in bistochastic classical theories can be obtained from their quantum counterparts by fixing an orthonormal basis for every local Hilbert space and setting all the off-diagonal elements of the Choi operator $S$ to zero. In the bipartite case, an admissible supermap is described by a matrix with entries $\sigma_{\st s    \st s'}   =   \< s_1|  \<  s_1'|  \<s_2| s_2'| S |s_1\>  |s_1'\>  |s_2\>  |s_2'\>$ where, for $j \in  \{1,2\}$,  $\{  |s_j\>\}$ and  $\{  |s_j'\>\}$ are bases for the Hilbert spaces of the input and output spaces of the $j$-th party.  
 % For this matrix, the analogue of Eq. (\ref{constraints}) is 
 %   \begin{align}\label{constraints2}
 % {\color{red}add~constraints~here}
 %  \end{align}
In particular,  the supermap described in the main text corresponds to the matrix  with entries $\sigma_{\st s    \st s'}   =  \delta_{s_1, s_1' \oplus s_2'}  \delta_{s_2, s_1' \oplus s_2'}$, which satisfies all the constraints in Eq. (\ref{constraints}) and therefore represents a logically admissible process in bistochastic quantum and classical theory.

\section*{Data availability}
The authors declare that the data supporting the findings of this study are available within the paper and in the supplementary information files.

\bibliography{references.bib}

\begin{thebibliography}{62}%
\makeatletter
\providecommand \@ifxundefined [1]{%
 \@ifx{#1\undefined}
}%
\providecommand \@ifnum [1]{%
 \ifnum #1\expandafter \@firstoftwo
 \else \expandafter \@secondoftwo
 \fi
}%
\providecommand \@ifx [1]{%
 \ifx #1\expandafter \@firstoftwo
 \else \expandafter \@secondoftwo
 \fi
}%
\providecommand \natexlab [1]{#1}%
\providecommand \enquote  [1]{``#1''}%
\providecommand \bibnamefont  [1]{#1}%
\providecommand \bibfnamefont [1]{#1}%
\providecommand \citenamefont [1]{#1}%
\providecommand \href@noop [0]{\@secondoftwo}%
\providecommand \href [0]{\begingroup \@sanitize@url \@href}%
\providecommand \@href[1]{\@@startlink{#1}\@@href}%
\providecommand \@@href[1]{\endgroup#1\@@endlink}%
\providecommand \@sanitize@url [0]{\catcode `\\12\catcode `\$12\catcode `\&12\catcode `\#12\catcode `\^12\catcode `\_12\catcode `\%12\relax}%
\providecommand \@@startlink[1]{}%
\providecommand \@@endlink[0]{}%
\providecommand \url  [0]{\begingroup\@sanitize@url \@url }%
\providecommand \@url [1]{\endgroup\@href {#1}{\urlprefix }}%
\providecommand \urlprefix  [0]{URL }%
\providecommand \Eprint [0]{\href }%
\providecommand \doibase [0]{http://dx.doi.org/}%
\providecommand \selectlanguage [0]{\@gobble}%
\providecommand \bibinfo  [0]{\@secondoftwo}%
\providecommand \bibfield  [0]{\@secondoftwo}%
\providecommand \translation [1]{[#1]}%
\providecommand \BibitemOpen [0]{}%
\providecommand \bibitemStop [0]{}%
\providecommand \bibitemNoStop [0]{.\EOS\space}%
\providecommand \EOS [0]{\spacefactor3000\relax}%
\providecommand \BibitemShut  [1]{\csname bibitem#1\endcsname}%
\let\auto@bib@innerbib\@empty
%</preamble>
\bibitem [{\citenamefont {Chiribella}\ \emph {et~al.}(2009{\natexlab{a}})\citenamefont {Chiribella}, \citenamefont {D’Ariano}, \citenamefont {Perinotti},\ and\ \citenamefont {Valiron}}]{chiribella2009beyond}%
  \BibitemOpen
  \bibfield  {author} {\bibinfo {author} {\bibfnamefont {G.}~\bibnamefont {Chiribella}}, \bibinfo {author} {\bibfnamefont {G.}~\bibnamefont {D’Ariano}}, \bibinfo {author} {\bibfnamefont {P.}~\bibnamefont {Perinotti}}, \ and\ \bibinfo {author} {\bibfnamefont {B.}~\bibnamefont {Valiron}},\ }\href@noop {} {\bibfield  {journal} {\bibinfo  {journal} {arXiv preprint arXiv:0912.0195}\ } (\bibinfo {year} {2009}{\natexlab{a}})}\BibitemShut {NoStop}%
\bibitem [{\citenamefont {Oreshkov}\ \emph {et~al.}(2012)\citenamefont {Oreshkov}, \citenamefont {Costa},\ and\ \citenamefont {Brukner}}]{oreshkov2012quantum}%
  \BibitemOpen
  \bibfield  {author} {\bibinfo {author} {\bibfnamefont {O.}~\bibnamefont {Oreshkov}}, \bibinfo {author} {\bibfnamefont {F.}~\bibnamefont {Costa}}, \ and\ \bibinfo {author} {\bibfnamefont {{\v{C}}.}~\bibnamefont {Brukner}},\ }\href@noop {} {\bibfield  {journal} {\bibinfo  {journal} {Nature Communications}\ }\textbf {\bibinfo {volume} {3}},\ \bibinfo {pages} {1} (\bibinfo {year} {2012})}\BibitemShut {NoStop}%
\bibitem [{\citenamefont {Chiribella}\ \emph {et~al.}(2013)\citenamefont {Chiribella}, \citenamefont {D{’}Ariano}, \citenamefont {Perinotti},\ and\ \citenamefont {Valiron}}]{chiribella2013quantum}%
  \BibitemOpen
  \bibfield  {author} {\bibinfo {author} {\bibfnamefont {G.}~\bibnamefont {Chiribella}}, \bibinfo {author} {\bibfnamefont {G.~M.}\ \bibnamefont {D{’}Ariano}}, \bibinfo {author} {\bibfnamefont {P.}~\bibnamefont {Perinotti}}, \ and\ \bibinfo {author} {\bibfnamefont {B.}~\bibnamefont {Valiron}},\ }\href@noop {} {\bibfield  {journal} {\bibinfo  {journal} {Physical Review A}\ }\textbf {\bibinfo {volume} {88}},\ \bibinfo {pages} {022318} (\bibinfo {year} {2013})}\BibitemShut {NoStop}%
\bibitem [{\citenamefont {Brukner}(2014)}]{brukner2014quantum}%
  \BibitemOpen
  \bibfield  {author} {\bibinfo {author} {\bibfnamefont {{\v{C}}.}~\bibnamefont {Brukner}},\ }\href@noop {} {\bibfield  {journal} {\bibinfo  {journal} {Nature Physics}\ }\textbf {\bibinfo {volume} {10}},\ \bibinfo {pages} {259} (\bibinfo {year} {2014})}\BibitemShut {NoStop}%
\bibitem [{\citenamefont {Chiribella}(2012)}]{chiribella2012perfect}%
  \BibitemOpen
  \bibfield  {author} {\bibinfo {author} {\bibfnamefont {G.}~\bibnamefont {Chiribella}},\ }\href@noop {} {\bibfield  {journal} {\bibinfo  {journal} {Physical Review A}\ }\textbf {\bibinfo {volume} {86}},\ \bibinfo {pages} {040301} (\bibinfo {year} {2012})}\BibitemShut {NoStop}%
\bibitem [{\citenamefont {Ara{\'u}jo}\ \emph {et~al.}(2014)\citenamefont {Ara{\'u}jo}, \citenamefont {Costa},\ and\ \citenamefont {Brukner}}]{araujo2014computational}%
  \BibitemOpen
  \bibfield  {author} {\bibinfo {author} {\bibfnamefont {M.}~\bibnamefont {Ara{\'u}jo}}, \bibinfo {author} {\bibfnamefont {F.}~\bibnamefont {Costa}}, \ and\ \bibinfo {author} {\bibfnamefont {{\v{C}}.}~\bibnamefont {Brukner}},\ }\href@noop {} {\bibfield  {journal} {\bibinfo  {journal} {Physical Review Letters}\ }\textbf {\bibinfo {volume} {113}},\ \bibinfo {pages} {250402} (\bibinfo {year} {2014})}\BibitemShut {NoStop}%
\bibitem [{\citenamefont {Gu{\'e}rin}\ \emph {et~al.}(2016)\citenamefont {Gu{\'e}rin}, \citenamefont {Feix}, \citenamefont {Ara{\'u}jo},\ and\ \citenamefont {Brukner}}]{guerin2016exponential}%
  \BibitemOpen
  \bibfield  {author} {\bibinfo {author} {\bibfnamefont {P.~A.}\ \bibnamefont {Gu{\'e}rin}}, \bibinfo {author} {\bibfnamefont {A.}~\bibnamefont {Feix}}, \bibinfo {author} {\bibfnamefont {M.}~\bibnamefont {Ara{\'u}jo}}, \ and\ \bibinfo {author} {\bibfnamefont {{\v{C}}.}~\bibnamefont {Brukner}},\ }\href@noop {} {\bibfield  {journal} {\bibinfo  {journal} {Physical Review Letters}\ }\textbf {\bibinfo {volume} {117}},\ \bibinfo {pages} {100502} (\bibinfo {year} {2016})}\BibitemShut {NoStop}%
\bibitem [{\citenamefont {Ebler}\ \emph {et~al.}(2018)\citenamefont {Ebler}, \citenamefont {Salek},\ and\ \citenamefont {Chiribella}}]{ebler2018enhanced}%
  \BibitemOpen
  \bibfield  {author} {\bibinfo {author} {\bibfnamefont {D.}~\bibnamefont {Ebler}}, \bibinfo {author} {\bibfnamefont {S.}~\bibnamefont {Salek}}, \ and\ \bibinfo {author} {\bibfnamefont {G.}~\bibnamefont {Chiribella}},\ }\href@noop {} {\bibfield  {journal} {\bibinfo  {journal} {Physical Review Letters}\ }\textbf {\bibinfo {volume} {120}},\ \bibinfo {pages} {120502} (\bibinfo {year} {2018})}\BibitemShut {NoStop}%
\bibitem [{\citenamefont {Zhao}\ \emph {et~al.}(2020)\citenamefont {Zhao}, \citenamefont {Yang},\ and\ \citenamefont {Chiribella}}]{zhao2020quantum}%
  \BibitemOpen
  \bibfield  {author} {\bibinfo {author} {\bibfnamefont {X.}~\bibnamefont {Zhao}}, \bibinfo {author} {\bibfnamefont {Y.}~\bibnamefont {Yang}}, \ and\ \bibinfo {author} {\bibfnamefont {G.}~\bibnamefont {Chiribella}},\ }\href@noop {} {\bibfield  {journal} {\bibinfo  {journal} {Physical Review Letters}\ }\textbf {\bibinfo {volume} {124}},\ \bibinfo {pages} {190503} (\bibinfo {year} {2020})}\BibitemShut {NoStop}%
\bibitem [{\citenamefont {Felce}\ and\ \citenamefont {Vedral}(2020)}]{felce2020quantum}%
  \BibitemOpen
  \bibfield  {author} {\bibinfo {author} {\bibfnamefont {D.}~\bibnamefont {Felce}}\ and\ \bibinfo {author} {\bibfnamefont {V.}~\bibnamefont {Vedral}},\ }\href@noop {} {\bibfield  {journal} {\bibinfo  {journal} {Physical Review Letters}\ }\textbf {\bibinfo {volume} {125}},\ \bibinfo {pages} {070603} (\bibinfo {year} {2020})}\BibitemShut {NoStop}%
\bibitem [{\citenamefont {Gao}\ \emph {et~al.}(2023)\citenamefont {Gao}, \citenamefont {Li}, \citenamefont {Mishra}, \citenamefont {Yan}, \citenamefont {Simonov},\ and\ \citenamefont {Chiribella}}]{gao2023measuring}%
  \BibitemOpen
  \bibfield  {author} {\bibinfo {author} {\bibfnamefont {N.}~\bibnamefont {Gao}}, \bibinfo {author} {\bibfnamefont {D.}~\bibnamefont {Li}}, \bibinfo {author} {\bibfnamefont {A.}~\bibnamefont {Mishra}}, \bibinfo {author} {\bibfnamefont {J.}~\bibnamefont {Yan}}, \bibinfo {author} {\bibfnamefont {K.}~\bibnamefont {Simonov}}, \ and\ \bibinfo {author} {\bibfnamefont {G.}~\bibnamefont {Chiribella}},\ }\href@noop {} {\bibfield  {journal} {\bibinfo  {journal} {Physical Review Letters}\ }\textbf {\bibinfo {volume} {130}},\ \bibinfo {pages} {170201} (\bibinfo {year} {2023})}\BibitemShut {NoStop}%
\bibitem [{\citenamefont {Zhu}\ \emph {et~al.}(2023)\citenamefont {Zhu}, \citenamefont {Chen}, \citenamefont {Hasegawa},\ and\ \citenamefont {Xue}}]{zhu2023charging}%
  \BibitemOpen
  \bibfield  {author} {\bibinfo {author} {\bibfnamefont {G.}~\bibnamefont {Zhu}}, \bibinfo {author} {\bibfnamefont {Y.}~\bibnamefont {Chen}}, \bibinfo {author} {\bibfnamefont {Y.}~\bibnamefont {Hasegawa}}, \ and\ \bibinfo {author} {\bibfnamefont {P.}~\bibnamefont {Xue}},\ }\href@noop {} {\bibfield  {journal} {\bibinfo  {journal} {Physical Review Letters}\ }\textbf {\bibinfo {volume} {131}},\ \bibinfo {pages} {240401} (\bibinfo {year} {2023})}\BibitemShut {NoStop}%
\bibitem [{\citenamefont {Hardy}(2007)}]{hardy2007towards}%
  \BibitemOpen
  \bibfield  {author} {\bibinfo {author} {\bibfnamefont {L.}~\bibnamefont {Hardy}},\ }\href@noop {} {\bibfield  {journal} {\bibinfo  {journal} {Journal of Physics A: Mathematical and Theoretical}\ }\textbf {\bibinfo {volume} {40}},\ \bibinfo {pages} {3081} (\bibinfo {year} {2007})}\BibitemShut {NoStop}%
\bibitem [{\citenamefont {Oreshkov}\ and\ \citenamefont {Giarmatzi}(2016)}]{oreshkov2016causal}%
  \BibitemOpen
  \bibfield  {author} {\bibinfo {author} {\bibfnamefont {O.}~\bibnamefont {Oreshkov}}\ and\ \bibinfo {author} {\bibfnamefont {C.}~\bibnamefont {Giarmatzi}},\ }\href@noop {} {\bibfield  {journal} {\bibinfo  {journal} {New Journal of Physics}\ }\textbf {\bibinfo {volume} {18}},\ \bibinfo {pages} {093020} (\bibinfo {year} {2016})}\BibitemShut {NoStop}%
\bibitem [{\citenamefont {Branciard}\ \emph {et~al.}(2015)\citenamefont {Branciard}, \citenamefont {Ara{\'u}jo}, \citenamefont {Feix}, \citenamefont {Costa},\ and\ \citenamefont {Brukner}}]{branciard2015simplest}%
  \BibitemOpen
  \bibfield  {author} {\bibinfo {author} {\bibfnamefont {C.}~\bibnamefont {Branciard}}, \bibinfo {author} {\bibfnamefont {M.}~\bibnamefont {Ara{\'u}jo}}, \bibinfo {author} {\bibfnamefont {A.}~\bibnamefont {Feix}}, \bibinfo {author} {\bibfnamefont {F.}~\bibnamefont {Costa}}, \ and\ \bibinfo {author} {\bibfnamefont {{\v{C}}.}~\bibnamefont {Brukner}},\ }\href@noop {} {\bibfield  {journal} {\bibinfo  {journal} {New Journal of Physics}\ }\textbf {\bibinfo {volume} {18}},\ \bibinfo {pages} {013008} (\bibinfo {year} {2015})}\BibitemShut {NoStop}%
\bibitem [{\citenamefont {Baumeler}\ \emph {et~al.}(2014)\citenamefont {Baumeler}, \citenamefont {Feix},\ and\ \citenamefont {Wolf}}]{baumeler2014maximal}%
  \BibitemOpen
  \bibfield  {author} {\bibinfo {author} {\bibfnamefont {{\"A}.}~\bibnamefont {Baumeler}}, \bibinfo {author} {\bibfnamefont {A.}~\bibnamefont {Feix}}, \ and\ \bibinfo {author} {\bibfnamefont {S.}~\bibnamefont {Wolf}},\ }\href@noop {} {\bibfield  {journal} {\bibinfo  {journal} {Physical Review A}\ }\textbf {\bibinfo {volume} {90}},\ \bibinfo {pages} {042106} (\bibinfo {year} {2014})}\BibitemShut {NoStop}%
\bibitem [{\citenamefont {Feix}\ \emph {et~al.}(2016)\citenamefont {Feix}, \citenamefont {Ara{\'u}jo},\ and\ \citenamefont {Brukner}}]{feix2016causally}%
  \BibitemOpen
  \bibfield  {author} {\bibinfo {author} {\bibfnamefont {A.}~\bibnamefont {Feix}}, \bibinfo {author} {\bibfnamefont {M.}~\bibnamefont {Ara{\'u}jo}}, \ and\ \bibinfo {author} {\bibfnamefont {{\v{C}}.}~\bibnamefont {Brukner}},\ }\href@noop {} {\bibfield  {journal} {\bibinfo  {journal} {New Journal of Physics}\ }\textbf {\bibinfo {volume} {18}},\ \bibinfo {pages} {083040} (\bibinfo {year} {2016})}\BibitemShut {NoStop}%
\bibitem [{\citenamefont {Abbott}\ \emph {et~al.}(2016)\citenamefont {Abbott}, \citenamefont {Giarmatzi}, \citenamefont {Costa},\ and\ \citenamefont {Branciard}}]{abbott2016multipartite}%
  \BibitemOpen
  \bibfield  {author} {\bibinfo {author} {\bibfnamefont {A.~A.}\ \bibnamefont {Abbott}}, \bibinfo {author} {\bibfnamefont {C.}~\bibnamefont {Giarmatzi}}, \bibinfo {author} {\bibfnamefont {F.}~\bibnamefont {Costa}}, \ and\ \bibinfo {author} {\bibfnamefont {C.}~\bibnamefont {Branciard}},\ }\href@noop {} {\bibfield  {journal} {\bibinfo  {journal} {Physical Review A}\ }\textbf {\bibinfo {volume} {94}},\ \bibinfo {pages} {032131} (\bibinfo {year} {2016})}\BibitemShut {NoStop}%
\bibitem [{\citenamefont {Wechs}\ \emph {et~al.}(2023)\citenamefont {Wechs}, \citenamefont {Branciard},\ and\ \citenamefont {Oreshkov}}]{wechs2023existence}%
  \BibitemOpen
  \bibfield  {author} {\bibinfo {author} {\bibfnamefont {J.}~\bibnamefont {Wechs}}, \bibinfo {author} {\bibfnamefont {C.}~\bibnamefont {Branciard}}, \ and\ \bibinfo {author} {\bibfnamefont {O.}~\bibnamefont {Oreshkov}},\ }\href@noop {} {\bibfield  {journal} {\bibinfo  {journal} {Nature Communications}\ }\textbf {\bibinfo {volume} {14}},\ \bibinfo {pages} {1471} (\bibinfo {year} {2023})}\BibitemShut {NoStop}%
\bibitem [{\citenamefont {Cirel'son}(1980)}]{cirel1980quantum}%
  \BibitemOpen
  \bibfield  {author} {\bibinfo {author} {\bibfnamefont {B.~S.}\ \bibnamefont {Cirel'son}},\ }\href@noop {} {\bibfield  {journal} {\bibinfo  {journal} {Letters in Mathematical Physics}\ }\textbf {\bibinfo {volume} {4}},\ \bibinfo {pages} {93} (\bibinfo {year} {1980})}\BibitemShut {NoStop}%
\bibitem [{\citenamefont {Popescu}\ and\ \citenamefont {Rohrlich}(1994)}]{popescu1994quantum}%
  \BibitemOpen
  \bibfield  {author} {\bibinfo {author} {\bibfnamefont {S.}~\bibnamefont {Popescu}}\ and\ \bibinfo {author} {\bibfnamefont {D.}~\bibnamefont {Rohrlich}},\ }\href@noop {} {\bibfield  {journal} {\bibinfo  {journal} {Foundations of Physics}\ }\textbf {\bibinfo {volume} {24}},\ \bibinfo {pages} {379} (\bibinfo {year} {1994})}\BibitemShut {NoStop}%
\bibitem [{\citenamefont {Van~Dam}(1999)}]{van1999nonlocality}%
  \BibitemOpen
  \bibfield  {author} {\bibinfo {author} {\bibfnamefont {W.}~\bibnamefont {Van~Dam}},\ }\emph {\bibinfo {title} {Nonlocality and communication complexity}},\ \href@noop {} {Ph.D. thesis},\ \bibinfo  {school} {University of Oxford} (\bibinfo {year} {1999})\BibitemShut {NoStop}%
\bibitem [{\citenamefont {Brassard}\ \emph {et~al.}(2006)\citenamefont {Brassard}, \citenamefont {Buhrman}, \citenamefont {Linden}, \citenamefont {M{\'e}thot}, \citenamefont {Tapp},\ and\ \citenamefont {Unger}}]{brassard2006limit}%
  \BibitemOpen
  \bibfield  {author} {\bibinfo {author} {\bibfnamefont {G.}~\bibnamefont {Brassard}}, \bibinfo {author} {\bibfnamefont {H.}~\bibnamefont {Buhrman}}, \bibinfo {author} {\bibfnamefont {N.}~\bibnamefont {Linden}}, \bibinfo {author} {\bibfnamefont {A.~A.}\ \bibnamefont {M{\'e}thot}}, \bibinfo {author} {\bibfnamefont {A.}~\bibnamefont {Tapp}}, \ and\ \bibinfo {author} {\bibfnamefont {F.}~\bibnamefont {Unger}},\ }\href@noop {} {\bibfield  {journal} {\bibinfo  {journal} {Physical Review Letters}\ }\textbf {\bibinfo {volume} {96}},\ \bibinfo {pages} {250401} (\bibinfo {year} {2006})}\BibitemShut {NoStop}%
\bibitem [{\citenamefont {Navascu{\'e}s}\ \emph {et~al.}(2007)\citenamefont {Navascu{\'e}s}, \citenamefont {Pironio},\ and\ \citenamefont {Ac{\'\i}n}}]{navascues2007bounding}%
  \BibitemOpen
  \bibfield  {author} {\bibinfo {author} {\bibfnamefont {M.}~\bibnamefont {Navascu{\'e}s}}, \bibinfo {author} {\bibfnamefont {S.}~\bibnamefont {Pironio}}, \ and\ \bibinfo {author} {\bibfnamefont {A.}~\bibnamefont {Ac{\'\i}n}},\ }\href@noop {} {\bibfield  {journal} {\bibinfo  {journal} {Physical Review Letters}\ }\textbf {\bibinfo {volume} {98}},\ \bibinfo {pages} {010401} (\bibinfo {year} {2007})}\BibitemShut {NoStop}%
\bibitem [{\citenamefont {Brunner}\ and\ \citenamefont {Skrzypczyk}(2009)}]{brunner2009nonlocality}%
  \BibitemOpen
  \bibfield  {author} {\bibinfo {author} {\bibfnamefont {N.}~\bibnamefont {Brunner}}\ and\ \bibinfo {author} {\bibfnamefont {P.}~\bibnamefont {Skrzypczyk}},\ }\href@noop {} {\bibfield  {journal} {\bibinfo  {journal} {Physical Review Letters}\ }\textbf {\bibinfo {volume} {102}},\ \bibinfo {pages} {160403} (\bibinfo {year} {2009})}\BibitemShut {NoStop}%
\bibitem [{\citenamefont {Linden}\ \emph {et~al.}(2007)\citenamefont {Linden}, \citenamefont {Popescu}, \citenamefont {Short},\ and\ \citenamefont {Winter}}]{linden2007quantum}%
  \BibitemOpen
  \bibfield  {author} {\bibinfo {author} {\bibfnamefont {N.}~\bibnamefont {Linden}}, \bibinfo {author} {\bibfnamefont {S.}~\bibnamefont {Popescu}}, \bibinfo {author} {\bibfnamefont {A.~J.}\ \bibnamefont {Short}}, \ and\ \bibinfo {author} {\bibfnamefont {A.}~\bibnamefont {Winter}},\ }\href@noop {} {\bibfield  {journal} {\bibinfo  {journal} {Physical Review Letters}\ }\textbf {\bibinfo {volume} {99}},\ \bibinfo {pages} {180502} (\bibinfo {year} {2007})}\BibitemShut {NoStop}%
\bibitem [{\citenamefont {Paw{\l}owski}\ \emph {et~al.}(2009)\citenamefont {Paw{\l}owski}, \citenamefont {Paterek}, \citenamefont {Kaszlikowski}, \citenamefont {Scarani}, \citenamefont {Winter},\ and\ \citenamefont {{\.Z}ukowski}}]{pawlowski2009information}%
  \BibitemOpen
  \bibfield  {author} {\bibinfo {author} {\bibfnamefont {M.}~\bibnamefont {Paw{\l}owski}}, \bibinfo {author} {\bibfnamefont {T.}~\bibnamefont {Paterek}}, \bibinfo {author} {\bibfnamefont {D.}~\bibnamefont {Kaszlikowski}}, \bibinfo {author} {\bibfnamefont {V.}~\bibnamefont {Scarani}}, \bibinfo {author} {\bibfnamefont {A.}~\bibnamefont {Winter}}, \ and\ \bibinfo {author} {\bibfnamefont {M.}~\bibnamefont {{\.Z}ukowski}},\ }\href@noop {} {\bibfield  {journal} {\bibinfo  {journal} {Nature}\ }\textbf {\bibinfo {volume} {461}},\ \bibinfo {pages} {1101} (\bibinfo {year} {2009})}\BibitemShut {NoStop}%
\bibitem [{\citenamefont {Navascu{\'e}s}\ and\ \citenamefont {Wunderlich}(2010)}]{navascues2010glance}%
  \BibitemOpen
  \bibfield  {author} {\bibinfo {author} {\bibfnamefont {M.}~\bibnamefont {Navascu{\'e}s}}\ and\ \bibinfo {author} {\bibfnamefont {H.}~\bibnamefont {Wunderlich}},\ }\href@noop {} {\bibfield  {journal} {\bibinfo  {journal} {Proceedings of the Royal Society A: Mathematical, Physical and Engineering Sciences}\ }\textbf {\bibinfo {volume} {466}},\ \bibinfo {pages} {881} (\bibinfo {year} {2010})}\BibitemShut {NoStop}%
\bibitem [{\citenamefont {Fritz}\ \emph {et~al.}(2013)\citenamefont {Fritz}, \citenamefont {Sainz}, \citenamefont {Augusiak}, \citenamefont {Brask}, \citenamefont {Chaves}, \citenamefont {Leverrier},\ and\ \citenamefont {Ac{\'\i}n}}]{fritz2013local}%
  \BibitemOpen
  \bibfield  {author} {\bibinfo {author} {\bibfnamefont {T.}~\bibnamefont {Fritz}}, \bibinfo {author} {\bibfnamefont {A.~B.}\ \bibnamefont {Sainz}}, \bibinfo {author} {\bibfnamefont {R.}~\bibnamefont {Augusiak}}, \bibinfo {author} {\bibfnamefont {J.~B.}\ \bibnamefont {Brask}}, \bibinfo {author} {\bibfnamefont {R.}~\bibnamefont {Chaves}}, \bibinfo {author} {\bibfnamefont {A.}~\bibnamefont {Leverrier}}, \ and\ \bibinfo {author} {\bibfnamefont {A.}~\bibnamefont {Ac{\'\i}n}},\ }\href@noop {} {\bibfield  {journal} {\bibinfo  {journal} {Nature Communications}\ }\textbf {\bibinfo {volume} {4}},\ \bibinfo {pages} {2263} (\bibinfo {year} {2013})}\BibitemShut {NoStop}%
\bibitem [{\citenamefont {Clauser}\ \emph {et~al.}(1969)\citenamefont {Clauser}, \citenamefont {Horne}, \citenamefont {Shimony},\ and\ \citenamefont {Holt}}]{clauser1969proposed}%
  \BibitemOpen
  \bibfield  {author} {\bibinfo {author} {\bibfnamefont {J.~F.}\ \bibnamefont {Clauser}}, \bibinfo {author} {\bibfnamefont {M.~A.}\ \bibnamefont {Horne}}, \bibinfo {author} {\bibfnamefont {A.}~\bibnamefont {Shimony}}, \ and\ \bibinfo {author} {\bibfnamefont {R.~A.}\ \bibnamefont {Holt}},\ }\href@noop {} {\bibfield  {journal} {\bibinfo  {journal} {Physical Review Letters}\ }\textbf {\bibinfo {volume} {23}},\ \bibinfo {pages} {880} (\bibinfo {year} {1969})}\BibitemShut {NoStop}%
\bibitem [{\citenamefont {Brukner}(2015)}]{brukner2015bounding}%
  \BibitemOpen
  \bibfield  {author} {\bibinfo {author} {\bibfnamefont {{\v{C}}.}~\bibnamefont {Brukner}},\ }\href@noop {} {\bibfield  {journal} {\bibinfo  {journal} {New Journal of Physics}\ }\textbf {\bibinfo {volume} {17}},\ \bibinfo {pages} {083034} (\bibinfo {year} {2015})}\BibitemShut {NoStop}%
\bibitem [{\citenamefont {Almeida}\ \emph {et~al.}(2010)\citenamefont {Almeida}, \citenamefont {Bancal}, \citenamefont {Brunner}, \citenamefont {Ac{\'\i}n}, \citenamefont {Gisin},\ and\ \citenamefont {Pironio}}]{almeida2010guess}%
  \BibitemOpen
  \bibfield  {author} {\bibinfo {author} {\bibfnamefont {M.~L.}\ \bibnamefont {Almeida}}, \bibinfo {author} {\bibfnamefont {J.-D.}\ \bibnamefont {Bancal}}, \bibinfo {author} {\bibfnamefont {N.}~\bibnamefont {Brunner}}, \bibinfo {author} {\bibfnamefont {A.}~\bibnamefont {Ac{\'\i}n}}, \bibinfo {author} {\bibfnamefont {N.}~\bibnamefont {Gisin}}, \ and\ \bibinfo {author} {\bibfnamefont {S.}~\bibnamefont {Pironio}},\ }\href@noop {} {\bibfield  {journal} {\bibinfo  {journal} {Physical Review Letters}\ }\textbf {\bibinfo {volume} {104}},\ \bibinfo {pages} {230404} (\bibinfo {year} {2010})}\BibitemShut {NoStop}%
\bibitem [{\citenamefont {Kunjwal}\ and\ \citenamefont {Oreshkov}(2023)}]{kunjwal2023nonclassicality}%
  \BibitemOpen
  \bibfield  {author} {\bibinfo {author} {\bibfnamefont {R.}~\bibnamefont {Kunjwal}}\ and\ \bibinfo {author} {\bibfnamefont {O.}~\bibnamefont {Oreshkov}},\ }\href@noop {} {\bibfield  {journal} {\bibinfo  {journal} {arXiv preprint arXiv:2307.02565}\ } (\bibinfo {year} {2023})}\BibitemShut {NoStop}%
\bibitem [{\citenamefont {Ara{\'u}jo}\ \emph {et~al.}(2015)\citenamefont {Ara{\'u}jo}, \citenamefont {Branciard}, \citenamefont {Costa}, \citenamefont {Feix}, \citenamefont {Giarmatzi},\ and\ \citenamefont {Brukner}}]{araujo2015witnessing}%
  \BibitemOpen
  \bibfield  {author} {\bibinfo {author} {\bibfnamefont {M.}~\bibnamefont {Ara{\'u}jo}}, \bibinfo {author} {\bibfnamefont {C.}~\bibnamefont {Branciard}}, \bibinfo {author} {\bibfnamefont {F.}~\bibnamefont {Costa}}, \bibinfo {author} {\bibfnamefont {A.}~\bibnamefont {Feix}}, \bibinfo {author} {\bibfnamefont {C.}~\bibnamefont {Giarmatzi}}, \ and\ \bibinfo {author} {\bibfnamefont {{\v{C}}.}~\bibnamefont {Brukner}},\ }\href@noop {} {\bibfield  {journal} {\bibinfo  {journal} {New Journal of Physics}\ }\textbf {\bibinfo {volume} {17}},\ \bibinfo {pages} {102001} (\bibinfo {year} {2015})}\BibitemShut {NoStop}%
\bibitem [{\citenamefont {Chiribella}\ and\ \citenamefont {Ebler}(2016)}]{chiribella2016optimal}%
  \BibitemOpen
  \bibfield  {author} {\bibinfo {author} {\bibfnamefont {G.}~\bibnamefont {Chiribella}}\ and\ \bibinfo {author} {\bibfnamefont {D.}~\bibnamefont {Ebler}},\ }\href@noop {} {\bibfield  {journal} {\bibinfo  {journal} {New Journal of Physics}\ }\textbf {\bibinfo {volume} {18}},\ \bibinfo {pages} {093053} (\bibinfo {year} {2016})}\BibitemShut {NoStop}%
\bibitem [{\citenamefont {Bisio}\ and\ \citenamefont {Perinotti}(2019)}]{bisio2019theoretical}%
  \BibitemOpen
  \bibfield  {author} {\bibinfo {author} {\bibfnamefont {A.}~\bibnamefont {Bisio}}\ and\ \bibinfo {author} {\bibfnamefont {P.}~\bibnamefont {Perinotti}},\ }\href@noop {} {\bibfield  {journal} {\bibinfo  {journal} {Proceedings of the Royal Society A}\ }\textbf {\bibinfo {volume} {475}},\ \bibinfo {pages} {20180706} (\bibinfo {year} {2019})}\BibitemShut {NoStop}%
\bibitem [{\citenamefont {Oreshkov}(2019)}]{oreshkov2019time}%
  \BibitemOpen
  \bibfield  {author} {\bibinfo {author} {\bibfnamefont {O.}~\bibnamefont {Oreshkov}},\ }\href@noop {} {\bibfield  {journal} {\bibinfo  {journal} {Quantum}\ }\textbf {\bibinfo {volume} {3}},\ \bibinfo {pages} {206} (\bibinfo {year} {2019})}\BibitemShut {NoStop}%
\bibitem [{\citenamefont {Baumeler}\ and\ \citenamefont {Wolf}(2016)}]{baumeler2016space}%
  \BibitemOpen
  \bibfield  {author} {\bibinfo {author} {\bibfnamefont {{\"A}.}~\bibnamefont {Baumeler}}\ and\ \bibinfo {author} {\bibfnamefont {S.}~\bibnamefont {Wolf}},\ }\href@noop {} {\bibfield  {journal} {\bibinfo  {journal} {New Journal of Physics}\ }\textbf {\bibinfo {volume} {18}},\ \bibinfo {pages} {013036} (\bibinfo {year} {2016})}\BibitemShut {NoStop}%
\bibitem [{\citenamefont {Bhattacharya}\ and\ \citenamefont {Banik}(2015)}]{bhattacharya2015biased}%
  \BibitemOpen
  \bibfield  {author} {\bibinfo {author} {\bibfnamefont {S.~S.}\ \bibnamefont {Bhattacharya}}\ and\ \bibinfo {author} {\bibfnamefont {M.}~\bibnamefont {Banik}},\ }\href@noop {} {\bibfield  {journal} {\bibinfo  {journal} {arXiv preprint arXiv:1509.02721}\ } (\bibinfo {year} {2015})}\BibitemShut {NoStop}%
\bibitem [{\citenamefont {Brunner}\ \emph {et~al.}(2014)\citenamefont {Brunner}, \citenamefont {Cavalcanti}, \citenamefont {Pironio}, \citenamefont {Scarani},\ and\ \citenamefont {Wehner}}]{brunner2014bell}%
  \BibitemOpen
  \bibfield  {author} {\bibinfo {author} {\bibfnamefont {N.}~\bibnamefont {Brunner}}, \bibinfo {author} {\bibfnamefont {D.}~\bibnamefont {Cavalcanti}}, \bibinfo {author} {\bibfnamefont {S.}~\bibnamefont {Pironio}}, \bibinfo {author} {\bibfnamefont {V.}~\bibnamefont {Scarani}}, \ and\ \bibinfo {author} {\bibfnamefont {S.}~\bibnamefont {Wehner}},\ }\href@noop {} {\bibfield  {journal} {\bibinfo  {journal} {Reviews of Modern Physics}\ }\textbf {\bibinfo {volume} {86}},\ \bibinfo {pages} {419} (\bibinfo {year} {2014})}\BibitemShut {NoStop}%
\bibitem [{\citenamefont {Bavaresco}\ \emph {et~al.}(2019)\citenamefont {Bavaresco}, \citenamefont {Ara{\'u}jo}, \citenamefont {Brukner},\ and\ \citenamefont {Quintino}}]{bavaresco2019semi}%
  \BibitemOpen
  \bibfield  {author} {\bibinfo {author} {\bibfnamefont {J.}~\bibnamefont {Bavaresco}}, \bibinfo {author} {\bibfnamefont {M.}~\bibnamefont {Ara{\'u}jo}}, \bibinfo {author} {\bibfnamefont {{\v{C}}.}~\bibnamefont {Brukner}}, \ and\ \bibinfo {author} {\bibfnamefont {M.~T.}\ \bibnamefont {Quintino}},\ }\href@noop {} {\bibfield  {journal} {\bibinfo  {journal} {Quantum}\ }\textbf {\bibinfo {volume} {3}},\ \bibinfo {pages} {176} (\bibinfo {year} {2019})}\BibitemShut {NoStop}%
\bibitem [{\citenamefont {Gross}\ \emph {et~al.}(2010)\citenamefont {Gross}, \citenamefont {M{\"u}ller}, \citenamefont {Colbeck},\ and\ \citenamefont {Dahlsten}}]{gross2010all}%
  \BibitemOpen
  \bibfield  {author} {\bibinfo {author} {\bibfnamefont {D.}~\bibnamefont {Gross}}, \bibinfo {author} {\bibfnamefont {M.}~\bibnamefont {M{\"u}ller}}, \bibinfo {author} {\bibfnamefont {R.}~\bibnamefont {Colbeck}}, \ and\ \bibinfo {author} {\bibfnamefont {O.~C.}\ \bibnamefont {Dahlsten}},\ }\href@noop {} {\bibfield  {journal} {\bibinfo  {journal} {Physical review letters}\ }\textbf {\bibinfo {volume} {104}},\ \bibinfo {pages} {080402} (\bibinfo {year} {2010})}\BibitemShut {NoStop}%
\bibitem [{\citenamefont {Chiribella}\ \emph {et~al.}(2021)\citenamefont {Chiribella}, \citenamefont {Aurell},\ and\ \citenamefont {{\.Z}yczkowski}}]{chiribella2021symmetries}%
  \BibitemOpen
  \bibfield  {author} {\bibinfo {author} {\bibfnamefont {G.}~\bibnamefont {Chiribella}}, \bibinfo {author} {\bibfnamefont {E.}~\bibnamefont {Aurell}}, \ and\ \bibinfo {author} {\bibfnamefont {K.}~\bibnamefont {{\.Z}yczkowski}},\ }\href@noop {} {\bibfield  {journal} {\bibinfo  {journal} {Physical Review Research}\ }\textbf {\bibinfo {volume} {3}},\ \bibinfo {pages} {033028} (\bibinfo {year} {2021})}\BibitemShut {NoStop}%
\bibitem [{\citenamefont {Chiribella}\ and\ \citenamefont {Liu}(2022)}]{chiribella2022quantum}%
  \BibitemOpen
  \bibfield  {author} {\bibinfo {author} {\bibfnamefont {G.}~\bibnamefont {Chiribella}}\ and\ \bibinfo {author} {\bibfnamefont {Z.}~\bibnamefont {Liu}},\ }\href@noop {} {\bibfield  {journal} {\bibinfo  {journal} {Communications Physics}\ }\textbf {\bibinfo {volume} {5}},\ \bibinfo {pages} {1} (\bibinfo {year} {2022})}\BibitemShut {NoStop}%
\bibitem [{\citenamefont {Navascu{\'e}s}\ \emph {et~al.}(2008)\citenamefont {Navascu{\'e}s}, \citenamefont {Pironio},\ and\ \citenamefont {Ac{\'\i}n}}]{navascues2008convergent}%
  \BibitemOpen
  \bibfield  {author} {\bibinfo {author} {\bibfnamefont {M.}~\bibnamefont {Navascu{\'e}s}}, \bibinfo {author} {\bibfnamefont {S.}~\bibnamefont {Pironio}}, \ and\ \bibinfo {author} {\bibfnamefont {A.}~\bibnamefont {Ac{\'\i}n}},\ }\href@noop {} {\bibfield  {journal} {\bibinfo  {journal} {New Journal of Physics}\ }\textbf {\bibinfo {volume} {10}},\ \bibinfo {pages} {073013} (\bibinfo {year} {2008})}\BibitemShut {NoStop}%
\bibitem [{\citenamefont {Gogioso}\ and\ \citenamefont {Pinzani}(2023)}]{gogioso2023geometry}%
  \BibitemOpen
  \bibfield  {author} {\bibinfo {author} {\bibfnamefont {S.}~\bibnamefont {Gogioso}}\ and\ \bibinfo {author} {\bibfnamefont {N.}~\bibnamefont {Pinzani}},\ }\href@noop {} {\bibfield  {journal} {\bibinfo  {journal} {arXiv preprint arXiv:2303.09017}\ } (\bibinfo {year} {2023})}\BibitemShut {NoStop}%
\bibitem [{\citenamefont {van~der Lugt}\ \emph {et~al.}(2023)\citenamefont {van~der Lugt}, \citenamefont {Barrett},\ and\ \citenamefont {Chiribella}}]{van2023device}%
  \BibitemOpen
  \bibfield  {author} {\bibinfo {author} {\bibfnamefont {T.}~\bibnamefont {van~der Lugt}}, \bibinfo {author} {\bibfnamefont {J.}~\bibnamefont {Barrett}}, \ and\ \bibinfo {author} {\bibfnamefont {G.}~\bibnamefont {Chiribella}},\ }\href@noop {} {\bibfield  {journal} {\bibinfo  {journal} {Nature Communications}\ }\textbf {\bibinfo {volume} {14}},\ \bibinfo {pages} {5811} (\bibinfo {year} {2023})}\BibitemShut {NoStop}%
\bibitem [{\citenamefont {Mayers}\ and\ \citenamefont {Yao}(1998)}]{mayers1998quantum}%
  \BibitemOpen
  \bibfield  {author} {\bibinfo {author} {\bibfnamefont {D.}~\bibnamefont {Mayers}}\ and\ \bibinfo {author} {\bibfnamefont {A.}~\bibnamefont {Yao}},\ }in\ \href@noop {} {\emph {\bibinfo {booktitle} {Proceedings 39th Annual Symposium on Foundations of Computer Science (Cat. No. 98CB36280)}}}\ (\bibinfo {organization} {IEEE},\ \bibinfo {year} {1998})\ pp.\ \bibinfo {pages} {503--509}\BibitemShut {NoStop}%
\bibitem [{\citenamefont {Mayers}\ and\ \citenamefont {Yao}(2004)}]{mayers2004self}%
  \BibitemOpen
  \bibfield  {author} {\bibinfo {author} {\bibfnamefont {D.}~\bibnamefont {Mayers}}\ and\ \bibinfo {author} {\bibfnamefont {A.}~\bibnamefont {Yao}},\ }\href@noop {} {\bibfield  {journal} {\bibinfo  {journal} {Quantum Info. Comput.}\ }\textbf {\bibinfo {volume} {4}},\ \bibinfo {pages} {273–286} (\bibinfo {year} {2004})}\BibitemShut {NoStop}%
\bibitem [{\citenamefont {Acin}\ \emph {et~al.}(2006)\citenamefont {Acin}, \citenamefont {Gisin},\ and\ \citenamefont {Masanes}}]{acin2006bell}%
  \BibitemOpen
  \bibfield  {author} {\bibinfo {author} {\bibfnamefont {A.}~\bibnamefont {Acin}}, \bibinfo {author} {\bibfnamefont {N.}~\bibnamefont {Gisin}}, \ and\ \bibinfo {author} {\bibfnamefont {L.}~\bibnamefont {Masanes}},\ }\href@noop {} {\bibfield  {journal} {\bibinfo  {journal} {Physical Review Letters}\ }\textbf {\bibinfo {volume} {97}},\ \bibinfo {pages} {120405} (\bibinfo {year} {2006})}\BibitemShut {NoStop}%
\bibitem [{\citenamefont {Colbeck}(2009)}]{colbeck2009quantum}%
  \BibitemOpen
  \bibfield  {author} {\bibinfo {author} {\bibfnamefont {R.}~\bibnamefont {Colbeck}},\ }\href@noop {} {\bibfield  {journal} {\bibinfo  {journal} {arXiv preprint arXiv:0911.3814}\ } (\bibinfo {year} {2009})}\BibitemShut {NoStop}%
\bibitem [{\citenamefont {Pironio}\ \emph {et~al.}(2010)\citenamefont {Pironio}, \citenamefont {Ac{\'\i}n}, \citenamefont {Massar}, \citenamefont {de~La~Giroday}, \citenamefont {Matsukevich}, \citenamefont {Maunz}, \citenamefont {Olmschenk}, \citenamefont {Hayes}, \citenamefont {Luo}, \citenamefont {Manning} \emph {et~al.}}]{pironio2010random}%
  \BibitemOpen
  \bibfield  {author} {\bibinfo {author} {\bibfnamefont {S.}~\bibnamefont {Pironio}}, \bibinfo {author} {\bibfnamefont {A.}~\bibnamefont {Ac{\'\i}n}}, \bibinfo {author} {\bibfnamefont {S.}~\bibnamefont {Massar}}, \bibinfo {author} {\bibfnamefont {A.~B.}\ \bibnamefont {de~La~Giroday}}, \bibinfo {author} {\bibfnamefont {D.~N.}\ \bibnamefont {Matsukevich}}, \bibinfo {author} {\bibfnamefont {P.}~\bibnamefont {Maunz}}, \bibinfo {author} {\bibfnamefont {S.}~\bibnamefont {Olmschenk}}, \bibinfo {author} {\bibfnamefont {D.}~\bibnamefont {Hayes}}, \bibinfo {author} {\bibfnamefont {L.}~\bibnamefont {Luo}}, \bibinfo {author} {\bibfnamefont {T.~A.}\ \bibnamefont {Manning}},  \emph {et~al.},\ }\href@noop {} {\bibfield  {journal} {\bibinfo  {journal} {Nature}\ }\textbf {\bibinfo {volume} {464}},\ \bibinfo {pages} {1021} (\bibinfo {year} {2010})}\BibitemShut {NoStop}%
\bibitem [{\citenamefont {Buhrman}\ \emph {et~al.}(2010)\citenamefont {Buhrman}, \citenamefont {Cleve}, \citenamefont {Massar},\ and\ \citenamefont {De~Wolf}}]{buhrman2010nonlocality}%
  \BibitemOpen
  \bibfield  {author} {\bibinfo {author} {\bibfnamefont {H.}~\bibnamefont {Buhrman}}, \bibinfo {author} {\bibfnamefont {R.}~\bibnamefont {Cleve}}, \bibinfo {author} {\bibfnamefont {S.}~\bibnamefont {Massar}}, \ and\ \bibinfo {author} {\bibfnamefont {R.}~\bibnamefont {De~Wolf}},\ }\href@noop {} {\bibfield  {journal} {\bibinfo  {journal} {Reviews of Modern Physics}\ }\textbf {\bibinfo {volume} {82}},\ \bibinfo {pages} {665} (\bibinfo {year} {2010})}\BibitemShut {NoStop}%
\bibitem [{\citenamefont {Reichardt}\ \emph {et~al.}(2013)\citenamefont {Reichardt}, \citenamefont {Unger},\ and\ \citenamefont {Vazirani}}]{reichardt2013classical}%
  \BibitemOpen
  \bibfield  {author} {\bibinfo {author} {\bibfnamefont {B.~W.}\ \bibnamefont {Reichardt}}, \bibinfo {author} {\bibfnamefont {F.}~\bibnamefont {Unger}}, \ and\ \bibinfo {author} {\bibfnamefont {U.}~\bibnamefont {Vazirani}},\ }\href@noop {} {\bibfield  {journal} {\bibinfo  {journal} {Nature}\ }\textbf {\bibinfo {volume} {496}},\ \bibinfo {pages} {456} (\bibinfo {year} {2013})}\BibitemShut {NoStop}%
\bibitem [{\citenamefont {Davies}\ and\ \citenamefont {Lewis}(1970)}]{davies1970operational}%
  \BibitemOpen
  \bibfield  {author} {\bibinfo {author} {\bibfnamefont {E.~B.}\ \bibnamefont {Davies}}\ and\ \bibinfo {author} {\bibfnamefont {J.~T.}\ \bibnamefont {Lewis}},\ }\href@noop {} {\bibfield  {journal} {\bibinfo  {journal} {Communications in Mathematical Physics}\ }\textbf {\bibinfo {volume} {17}},\ \bibinfo {pages} {239} (\bibinfo {year} {1970})}\BibitemShut {NoStop}%
\bibitem [{\citenamefont {Ozawa}(1984)}]{ozawa1984quantum}%
  \BibitemOpen
  \bibfield  {author} {\bibinfo {author} {\bibfnamefont {M.}~\bibnamefont {Ozawa}},\ }\href@noop {} {\bibfield  {journal} {\bibinfo  {journal} {Journal of Mathematical Physics}\ }\textbf {\bibinfo {volume} {25}},\ \bibinfo {pages} {79} (\bibinfo {year} {1984})}\BibitemShut {NoStop}%
\bibitem [{\citenamefont {Heinosaari}\ and\ \citenamefont {Ziman}(2011)}]{heinosaari2011mathematical}%
  \BibitemOpen
  \bibfield  {author} {\bibinfo {author} {\bibfnamefont {T.}~\bibnamefont {Heinosaari}}\ and\ \bibinfo {author} {\bibfnamefont {M.}~\bibnamefont {Ziman}},\ }\href@noop {} {\emph {\bibinfo {title} {The mathematical language of quantum theory: from uncertainty to entanglement}}}\ (\bibinfo  {publisher} {Cambridge University Press},\ \bibinfo {year} {2011})\BibitemShut {NoStop}%
\bibitem [{\citenamefont {Jamio{\l}kowski}(1972)}]{jamiolkowski1972linear}%
  \BibitemOpen
  \bibfield  {author} {\bibinfo {author} {\bibfnamefont {A.}~\bibnamefont {Jamio{\l}kowski}},\ }\href@noop {} {\bibfield  {journal} {\bibinfo  {journal} {Reports on Mathematical Physics}\ }\textbf {\bibinfo {volume} {3}},\ \bibinfo {pages} {275} (\bibinfo {year} {1972})}\BibitemShut {NoStop}%
\bibitem [{\citenamefont {Choi}(1975)}]{choi1975completely}%
  \BibitemOpen
  \bibfield  {author} {\bibinfo {author} {\bibfnamefont {M.-D.}\ \bibnamefont {Choi}},\ }\href@noop {} {\bibfield  {journal} {\bibinfo  {journal} {Linear algebra and its applications}\ }\textbf {\bibinfo {volume} {10}},\ \bibinfo {pages} {285} (\bibinfo {year} {1975})}\BibitemShut {NoStop}%
\bibitem [{\citenamefont {Chiribella}\ \emph {et~al.}(2008)\citenamefont {Chiribella}, \citenamefont {D'Ariano},\ and\ \citenamefont {Perinotti}}]{chiribella2008transforming}%
  \BibitemOpen
  \bibfield  {author} {\bibinfo {author} {\bibfnamefont {G.}~\bibnamefont {Chiribella}}, \bibinfo {author} {\bibfnamefont {G.~M.}\ \bibnamefont {D'Ariano}}, \ and\ \bibinfo {author} {\bibfnamefont {P.}~\bibnamefont {Perinotti}},\ }\href@noop {} {\bibfield  {journal} {\bibinfo  {journal} {EPL (Europhysics Letters)}\ }\textbf {\bibinfo {volume} {83}},\ \bibinfo {pages} {30004} (\bibinfo {year} {2008})}\BibitemShut {NoStop}%
\bibitem [{\citenamefont {Chiribella}\ \emph {et~al.}(2009{\natexlab{b}})\citenamefont {Chiribella}, \citenamefont {D’Ariano},\ and\ \citenamefont {Perinotti}}]{chiribella2009theoretical}%
  \BibitemOpen
  \bibfield  {author} {\bibinfo {author} {\bibfnamefont {G.}~\bibnamefont {Chiribella}}, \bibinfo {author} {\bibfnamefont {G.~M.}\ \bibnamefont {D’Ariano}}, \ and\ \bibinfo {author} {\bibfnamefont {P.}~\bibnamefont {Perinotti}},\ }\href@noop {} {\bibfield  {journal} {\bibinfo  {journal} {Physical Review A}\ }\textbf {\bibinfo {volume} {80}},\ \bibinfo {pages} {022339} (\bibinfo {year} {2009}{\natexlab{b}})}\BibitemShut {NoStop}%
\bibitem [{\citenamefont {Watrous}(2018)}]{watrous2018theory}%
  \BibitemOpen
  \bibfield  {author} {\bibinfo {author} {\bibfnamefont {J.}~\bibnamefont {Watrous}},\ }\href@noop {} {\emph {\bibinfo {title} {The theory of quantum information}}}\ (\bibinfo  {publisher} {Cambridge university press},\ \bibinfo {year} {2018})\BibitemShut {NoStop}%
\end{thebibliography}%

\section*{Acknowledgments}
The authors thank R. Ramanathan, \v C. Brukner, M. Araujo, O. Oreshokov, R. Kunjwal and S. Yoshida for helpful  comments.  This work has been supported by the Hong Kong Research Grant Council through the Senior Research Fellowship Scheme SRFS2021-7S02 and through the Research Impact Grant R7035-21F, and by the F.R.S.-FNRS under project CHEQS within the Excellence of Science (EOS) program.  The work was also supported  by the John Templeton Foundation through grant 62312, The Quantum Information Structure of Spacetime (qiss.fr).
The opinions expressed in this publication are those of the authors and do not necessarily reflect the views of the John Templeton Foundation. Research at the Perimeter Institute is supported by the Government of Canada through the Department of Innovation, Science and Economic Development Canada and by the Province of Ontario through the Ministry of Research, Innovation and Science.

\section*{Author contributions}
ZL introduced the notion of single-trigger correlations, derived the bounds on quantum ICO correlations, and showed the achievability of the algebraic maximum by processes in bistochastic  classical and quantum theories. GC supervised the research and proved the reduction to labelled projective measurements (Theorem 1).  GC and ZL wrote the manuscript. 

\medskip
\section*{Competing interests}
The authors declare no competing interests.

\appendix
\section{Semidefinite programming (SDP) relaxation of ICO bounds}
\label{app:sdp}

In this section, we introduce \CJ isomorphism which allows us to represent quantum processes and local quantum operations as operators; then we explicitly present the SDPs for computing the ICO bound of single-trigger correlations and for computing upper bounds of general correlations. 

\subsection*{\CJ isomorphism} 
Through \CJ isomorphism \cite{jamiolkowski1972linear,choi1975completely}, a linear map $\map K$ on operators can be conveniently represented as a Choi operator, defined as follows:
\begin{equation}
    \op{Choi}(\map K) := \sum_{i,j} |i\>\<j| \otimes \map K(|i\>\<j|) \, .
\end{equation}
If $\map K$ is a completely positive map with Kraus operators $\{ K_i \}$, it holds that
\begin{equation}
    \op{Choi}(\map K) = \sum_i |K_i\kk\bb K_i| \, ,
\end{equation}
where we use the double-ket notation defined as $|M\kk = \sum_j |j\> \otimes M|j\>$.

\subsection*{Quantum supermaps and process matrices} 
In an experiment without assumption of global causal structure, the local operations are described by quantum instruments, that is,  collections of linear maps $(  \map M_j)_j$ labelled by possible measurement outcomes $j$.   Each linear map  $\map M_j$, called a quantum operation,    transforms density matrices on an input system into (generally subnormalized) density matrices on an output system,   and is required to be completely positive.   In the following, we will denote the $i$-th party's instruments (conditional on the setting $x_i$) by $\left(\map M^{(i)}_{a_i \mid x_i}\right)_{a_i}$,  each map $\map M^{(i)}_{a_i \mid x_i}$  transforming quantum states of an input system $A_{\rm I}^{(i)}$  into quantum states of an output system $A_{\rm O}^{(i)}$. 
The most general rule for assigning probabilities to outcomes is provided by a process  $\map S$ that transforms all local quantum operations $(\map M^{(1)}, \cdots, \map M^{(N)})$ into probabilities $\map S  (\map M^{(1)}, \cdots, \map M^{(N)})$. In this way,  the joint conditional probability distribution of outcomes is given by 
\begin{align}
\label{eq:icoprob}
p  (\vec a \mid \vec x)   =   \map S   \left(  \map M^{(1)}_{a_1 \mid x_1}, \cdots, \map M^{(N)}_{a_N \mid x_N}    \right) \, .
\end{align}

Mathematically, a process is a completely positive (CP) map transforming a list of channels to the unit probability, which is a special case of quantum supermaps, the physically admissible operations transforming quantum channels \cite{chiribella2008transforming,chiribella2009theoretical}. Given a supermap $\map S$, we can induce a map $\widehat{\map S}$ which transforms the Choi operators of channels,
\begin{equation}
    \widehat{\map S}: \quad \op{Choi}(\map C) \mapsto \op{Choi}(\map S(\map C)) \, ,
\end{equation}
and we define the Choi operator of $\map S$ to be the Choi operator of $\widehat{\map S}$, i.e. $\op{Choi}(\map S) := \op{Choi}(\widehat{\map S})$. The condition that $\map S$ is a CP map is equivalent to the condition that $\op{Choi}(\map S)$ is a postive operator, i.e. $\op{Choi}(\map S) \geq 0$. 

The constraints on a process $\map S$ that are in principle compatible with the validity of quantum mechanics in all local laboratories can be completely   characterized in terms of the Choi operator $S$ of the process $\map S$, called the {\em process matrix}  \cite{oreshkov2012quantum}, satisfying the conditions (i) $S \geq 0$; (ii) $S \in \op{DualAff}(\op{Choi}(\set{NoSig}))$ where $\op{Choi}(\set{NoSig}))$ is the set of Choi operators of no-signalling channels and $\op{DualAff}$ denotes the dual affine space. See Refs. \cite{oreshkov2012quantum,araujo2015witnessing} for an explicit characterization of process matrices.
The joint conditional probability in Eq. (\ref{eq:icoprob}) is then computed by
\begin{equation}
    \label{eq:correlation_ico}
    p(\vec a \mid \vec x) = \Tr\left[ S^T \cdot \bigotimes_{i=1}^N M_{a_i \mid x_i}^{(i)} \right] \, .
\end{equation}
where $M_{a_i \mid x_i}^{(i)}$ are Choi operators of the corresponding instruments $\map M_{a_i \mid x_i}^{(i)}$ and $S^T$ represents transposition of the process matrix in the computational basis.

\subsection*{SDP duality}

In Methods, with fixed instruments $\left(\map M^{(i)}_{a_i \mid x_i}\right)_{a_i}$, we associate every correlation $\map I = \sum_{\vec a, \vec x} \alpha_{\vec a, \vec x}\, p(\vec a \mid \vec x)$ with a map $\map M_{\map I}$. The Choi operator of $\map M_{\map I}$ will be referred to as the performance operator, which is given by
\begin{equation}
    \Omega_{\map I} := \op{Choi}(\map M_{\map I}) = \sum_{\vec a, \vec x} \alpha_{\vec a,\vec x}\, \bigotimes_{i=1}^N \op{Choi}(\map M_{a_i \mid x_i}^{(i)}) \, ,
\end{equation}
In particular, when $\map I$ is a single-trigger correlation with triggers $\vec\xi$, we associated it with a single-trigger operator, which is the performance operator $\Omega_{\map I}^*$ with the canonical choice of instruments, and is in turn the Choi operator of the map $\map M_{\map I}^*$ in the main text. The single-trigger operator $\Omega_{\map I}^*$ is given by
\begin{equation}
    \Omega_{\map I}^* = \sum_{\vec a, \vec x} \alpha_{\vec a,\vec x}\, P_{\vec\xi,\vec a, \vec x}
\end{equation}
where $P_{\vec\xi,\vec a, \vec x}$ are projectors defined as
\begin{equation}
    P_{\vec\xi,\vec a, \vec x} := \bigotimes_{i=1}^N P_{\xi_i,a_i,x_i}^{(i)}
\end{equation}
with 
\begin{equation}
    P_{\xi_i,a_i,x_i}^{(i)} = \begin{cases}
        |a_i\>\<a_i| \otimes |a_i\>\<a_i| \otimes |\xi_i\>\<\xi_i| & x_i = \xi_i \, , \\
        |\Phi\>\< \Phi| \otimes |x_i\>\<x_i| & x_i \neq \xi_i \, ,
    \end{cases}
\end{equation}
where $|\Phi\> := \sum_j |j\> \otimes |j\> / \sqrt{m_i}$ is the canonical maximally entangled state.

With respective to a performance operator $\Omega_{\map I}$, the optimization of the corresponding correlation $\map I$ can be expressed through the following SDP \cite{chiribella2016optimal}:
\begin{alignat}{2}
    \label{eq:primal_prob}
    & \text{maximize} \quad && \Tr(\Omega_{\map I} S) \\
    & \text{subject to} \quad && S \in \op{DualAff}(\op{Choi}(\set{NoSig})) \nonumber \\
    & \quad && S \geq 0 \nonumber \, .
\end{alignat}
The dual problem of (\ref{eq:primal_prob}) is
\begin{alignat}{2}
    \label{eq:dual_prob}
    & \text{minimize} \quad && \eta \\
    & \text{subject to} \quad && \eta C \geq \Omega_{\map I} \nonumber \\
    & \quad && C \in \op{Aff}(\op{Choi}(\set{NoSig})) \nonumber \, ,
\end{alignat}
where $\op{Aff}(\op{Choi}(\set{NoSig}))$ denotes the set of affine combinations of the Choi operators of no-signalling channels.

By the property of duality, every dual feasible point $(\eta, C)$ provides an upper bound $\eta$ of the SDP pair (\ref{eq:primal_prob}, \ref{eq:dual_prob}). 
Furthermore, strong duality holds because both primal optimal value and dual optimal value are finite and the primal feasible set contains a strictly positive operator (the process matrix proportional to the identity) \cite{watrous2018theory,chiribella2016optimal}. 
In particular, the dual problem (\ref{eq:dual_prob}), with the performance operator being a single-trigger operator $\Omega_{\map I}^*$, is the SDP corresponding to the ICO bound of the corresponding single-trigger correlation $\map I$. When the coefficients of a single-trigger correlation $\map I$ are non-negative, the corresponding single-trigger operator $\Omega_{\map I}^*$ is positive and thus the dual problem (\ref{eq:dual_prob}) is equivalent to $\min_{C \in \op{Choi}(\set{NoSig})} \{ \eta \in \R \mid \Omega_{\map I}^* \leq \eta C  \} = 2^{D_{\max}(\Omega_{\map I}^* \| \op{Choi}(\set{NoSig}))}$, which is in turn equivalent to the formula $\map I^{\rm ICO} = 2^{D_{\max}(\map M_{\map I}^* \| \set{NoSig})}$ in the main text.

% We will denote the value of the SDP pair (\ref{eq:primal_prob}, \ref{eq:dual_prob}) as a convex function $\eta(\Omega_{\map I})$ of the performance operator $\Omega_{\map I}$. The function $\eta$ is equivalent to (via \CJ isomorphism) the function $\upsilon$ in the main text defined on the set of maps $\map M_{\map I}$.

\subsection*{SDP relaxation in an explicit form}

Now we present an SDP relaxation of the problem of computing the ICO bound in an explicit form. 
For given triggers $\vec\xi$, define the following operator for every setting vector $\vec x$ and outcome vector $\vec a$:
\begin{equation}
    Q_{\vec\xi,\vec a, \vec x} := \bigotimes_{i=1}^N Q_{\xi_i,a_i,x_i}^{(i)}
\end{equation}
with 
\begin{equation}
    Q_{\xi_i,a_i,x_i}^{(i)} = \begin{cases}
        |a_i\>\<a_i| \otimes |a_i\>\<a_i| \otimes |\xi_i\>\<\xi_i| & x_i = \xi_i \, , \\
        \frac {1} {m_i} |\Phi\>\< \Phi| \otimes |x_i\>\<x_i| & x_i \neq \xi_i \, .
    \end{cases}
\end{equation}

By the definition of single-trigger operator, a single-trigger operator $\Omega_{\vec\xi}$ with triggers $\vec\xi$ is a linear combination of the projectors $\left\{P_{\vec\xi, \vec a, \vec x} \right\}_{\vec a, \vec x}$, which is in turn equivalent to the following constraint:
\begin{align}
    \label{eq:operatorcons}
    \Omega_{\vec\xi} = \sum_{\vec a,\vec x} \Tr[\Omega_{\vec\xi}~Q_{\vec\xi,\vec a,\vec x}] P_{\vec\xi,\vec a,\vec x} \, ,
\end{align}
and the coefficients of the corresponding single-trigger correlation are given by
\begin{equation}
    \label{eq:recovercoeff}
    \alpha^{\vec\xi}_{\vec a,\vec x} = \Tr[\Omega_{\vec\xi}~Q_{\vec\xi,\vec a,\vec x}] \, .
\end{equation}

Combining Eq. (\ref{eq:operatorcons}), Eq. (\ref{eq:recovercoeff}) and the dual SDP (\ref{eq:dual_prob}) for single-trigger correlations, the optimization over the decomposition of a general correlation $\map I = \sum_{\vec a, \vec x} \alpha_{\vec a, \vec x}\, p(\vec a \mid \vec x)$ into single-trigger correlations can be formulated into the following SDP:
\begin{alignat}{2}
    \label{eq:sdpgeneral}
    & \text{minimize} \quad && \frac {\sum_{\vec\xi} \Tr(C_{\vec\xi})} {\prod_{i=1}^N m_i} \\
    & \text{subject to} \quad && \forall \vec\xi \quad \Omega_{\vec\xi} = \sum_{\vec a,\vec x} \Tr[\Omega_{\vec\xi}~Q_{\vec\xi,\vec a,\vec x}] P_{\vec\xi,\vec a,\vec x} \, , \nonumber \\
    & \quad && \forall \vec a~\forall \vec x \quad \alpha_{\vec a,\vec x} = \sum_{\vec\xi} \Tr[\Omega_{\vec\xi}~Q_{\vec\xi,\vec a,\vec x}] \nonumber \, , \\
    & \quad && \forall \vec\xi \quad C_{\vec\xi} \in \op{Span}(\op{Choi}(\set{NoSig})) \, , \nonumber \\
    & \quad && \forall \vec\xi \quad C_{\vec\xi} \geq \Omega_{\vec\xi} \, . \nonumber
\end{alignat}

\subsection*{Necessary conditions on the quantum ICO set}

From another point of view, the ICO bounds of single-trigger correlations are equivalent to a set of necessary conditions on the quantum ICO set. To characterize these necessary conditions, we introduce a projection operator, denoted as $\Pi_{\vec\xi}$, which acts on the space of conditional probability distributions. This projection discards the outcome $a_i$ and substitutes it with a random outcome for every party $i$ whose setting $x_i \neq \xi_i$. If a conditional probability distribution $(p(\vec a\mid \vec x))_{\vec a,\vec x}$ belongs to the quantum ICO set, then $\Pi_{\vec\xi}(p)$ can be realized by a process matrix along with the canonical instruments. This can be formulated through the following SDP:
\begin{alignat}{2}
    & \text{find} \quad && S \label{eq:necessarycond} \\
    & \text{subject to} \quad && \forall \vec a~\forall \vec x \quad \Pi_{\vec\xi}(p)(\vec a \mid \vec x) = \Tr(S P_{\vec\xi,\vec a,\vec x})  \nonumber \\
    & \quad && S \in \op{DualAff}(\op{Choi}(\set{NoSig})) \nonumber \\
    & \quad && S \geq 0 \nonumber \, .
\end{alignat}
We have in total $\prod_i n_i$ ($n_i$ is the number of settings for the $i$-th party) necessary conditions in the form of the SDP (\ref{eq:necessarycond}), corresponding to the choices of triggers.

\medskip
\section{Biased OCB correlations}
\label{app:ocblemma}

To show that the value $(1+\alpha + \sqrt{1+\alpha^2})/2$ is maximal for a biased OCB correlation, we decompose the biased OCB correlation into a random mixture of two single-trigger correlations, i.e.
\begin{align}
        &P(a_1 = b \mid c = 0) + \alpha\, P(a_2 = x_1 \mid c = 1) \nonumber \\
        & = \sum_{z\in \{0,1\}} \frac 1 2 \Big (P(a_1 = b \mid x_1 = z, c = 0) \nonumber \\
        &\qquad + \alpha\, P(a_2 = x_1 \mid b = z, c = 1) \Big) \label{eq:lazydecom} \, .
\end{align}
The proof is then complete with the following lemma, which indicates that both correlations in the decomposition is upper bounded by $(1+\alpha + \sqrt{1+\alpha^2})/2$.
\begin{lemma}
    \label{theo:ocblemma}
    For any $x_*, b_* \in \{0,1\}$ and $\alpha\in \R$, the ICO bound of the following single-trigger correlation
    \begin{align}
        &P(a_1 = b \mid x_1 = x_*, c = 0) \nonumber \\
        &+ \alpha\, P(a_2 = x_1 \mid b = b_*, c = 1) \label{eq:ocblazy}
    \end{align}
    is $\left( 1 + \alpha + \sqrt{1+\alpha^2} \right) / 2$.
\end{lemma}
\Proof
To derive an upper bound of the single-trigger correlation (\ref{eq:ocblazy}), we construct a feasible solution of the dual SDP (\ref{eq:dual_prob}). We claim that the following positive operator $C$ on a 7-qubit Hilbert space $(\spc H_1 \otimes \spc H_{2,3}) \otimes (\spc H_4 \otimes \spc H_{5,6,7})$ is equal to $\left((1+\alpha) + \sqrt{1+\alpha^2} \right)/2$ times the Choi operator of a  no-signalling channel
\begin{equation}
    C := \Omega + \sum_{k,l} |\Psi_{k,l}\>\<\Psi_{k,l}| \, ,
\end{equation}
where $\Omega$ is the single-trigger operator of the correlation (\ref{eq:ocblazy}):
\begin{equation}
    \begin{split}
        \Omega &:= 
        \frac 1 2 |00\>\<00| \otimes |x_*\>\< x_*| \otimes |I\kk\bb I| \otimes |0\>\<0| \otimes |0\>\<0| \\
        &\quad + \frac 1 2 |11\>\<11| \otimes |x_*\>\< x_*| \otimes |I\kk\bb I| \otimes |1\>\<1| \otimes |0\>\<0| \\
        &\quad + \frac \alpha 2 |I\kk\bb I| \otimes |0\>\< 0| \otimes |00\>\< 00| \otimes |b_*\>\<b_*| \otimes |1\>\<1| \\
        &\quad + \frac \alpha 2 |I\kk\bb I| \otimes |1\>\< 1| \otimes |11\>\< 11| \otimes |b_*\>\<b_*| \otimes |1\>\<1| \\
    \end{split}
\end{equation}
and
\begin{align}
    |\Psi_{k,l}\> = \sum_{i,j} c_{i,j,k,l}\, |i,i\> \otimes |k\> \otimes |j,j\> \otimes |l,0\> \, ,
\end{align}
with real coefficients $c_{i,j,k,l}$.
The conditions that $C$ is proportional to the Choi operator of a no-signalling channel are
\begin{equation}
    \Tr_{2,3} C = \frac{I_1} 2 \otimes \Tr_{1,2,3}C \, , \quad
    \Tr_{5,6,7} C = \frac{I_4} 2 \otimes \Tr_{4,5,6,7}C \, ,
\end{equation}
which are equivalent to the following equations:
\begin{align}
    \forall j,l \quad &\sum_k c_{1,j,k,l}^2 - c_{0,j,k,l}^2 = \frac{(-1)^l} 2 \, , \label{eq:nosigstart} \\
    \forall i,k \quad &\sum_l c_{i,1,k,l}^2 - c_{i,0,k,l}^2 = \frac{(-1)^k\alpha} 2 \, , \\
    \forall l \quad &\sum_k c_{1,0,k,l}\, c_{1,1,k,l} - c_{0,0,k,l}\, c_{0,1,k,l} = \frac{(-1)^l} 2 \, , \\
    \forall k \quad &\sum_l c_{0,1,k,l}\, c_{1,1,k,l} - c_{0,0,k,l}\, c_{1,0,k,l} = \frac{(-1)^k\alpha} 2 \, . \label{eq:nosigend}
\end{align}
We find the following solution of Eq. (\ref{eq:nosigstart})-(\ref{eq:nosigend})
\begin{equation}
    c_{i,j,k,l} = \frac {\left(1+\alpha^2 \right)^{1/4}} {4} \left(1 - \frac{(-1)^{i+l}}{\sqrt{1+\alpha^2}} - \frac{(-1)^{j+k}\alpha}{\sqrt{1+\alpha^2}} \right) \, ,
\end{equation}
with which $\Tr(C) = \Tr(\Omega) + \sum_{i,j,k,l} |c_{i,j,k,l}|^2 = 2(1+\alpha) + 2\sqrt{1+\alpha^2}$, which is $\left((1+\alpha) + \sqrt{1+\alpha^2} \right)/2$ times the trace of the Choi operator of a  no-signalling channel. Therefore, $\left((1+\alpha) + \sqrt{1+\alpha^2} \right)/2$ is an upper bound of the single-trigger correlation (\ref{eq:ocblazy}).

The bound can be attained with the instruments and a modification of the process matrix proposed in Ref. \cite{oreshkov2012quantum}. Suppose that Alice and Bob share the following process matrix
\begin{equation}
    \begin{split}
    S_{{\rm OCB},\alpha} &= \frac 1 4 I^{\otimes 4} + \frac {\alpha} {4\sqrt{1+\alpha^2}} I \otimes Z \otimes Z \otimes I \\
    &\quad + \frac {1} {4\sqrt{1+\alpha^2}} Z \otimes I \otimes X \otimes Z \, ,
    \end{split}
\end{equation}
where $X$ and $Z$ are Pauli-$X$ and Pauli-$Z$ matrices, respectively. To attain the bound, Alice measures her input qubit in computational basis and reprepare the her setting $x_1$ in the same basis, while Bob
\begin{itemize}
    \item (in the case $c=1$) measures his qubit in computational basis and reprepares an arbitrary qubit state;
    \item (in the case $c=0$) measures his qubit in Fourier basis and reprepare $a_2\oplus b$ in computation basis.
\end{itemize}
The Choi representation of Alice's and Bob's instruments are
\begin{equation}
    M_{a_1\mid x_1} = |a_1\>\<a_1| \otimes |x_1\>\<x_1| \, ,
\end{equation}
and
\begin{align}
    N_{a_2 \mid b, 0} &= |X_{a_2}\>\< X_{a_2}| \otimes |a_2 \oplus b\>\<a_2 \oplus b| \, , \\
    N_{a_2 \mid b, 1} &= |a_2\>\<a_2| \otimes \rho \, ,
\end{align}
where $\rho$ is an arbitrary density operator on a qubit and $|X_{a_2}\> = (|0\> + (-1)^{a_2} |1\>)/\sqrt 2$ are the eigenstates of Pauli-$X$ matrix (Fourier basis).
The joint conditional probability distribution between Alice's and Bob's outcomes $p(a_1,a_2\mid x_1,b,c) = \Tr \left[{S_{\rm OCB, \alpha}}^T (M_{a_1\mid x_1} \otimes N_{a_2 \mid b, c}) \right]$ is computed to be
\begin{align}
    p(a_1,a_2 \mid x_1,b,0) &= \frac 1 4 + \frac{(-1)^{a_1+b}}{4\sqrt{1+\alpha^2}} \, , \\
    p(a_1,a_2 \mid x_1,b,1) &= \frac 1 4 + \frac{(-1)^{a_2+x_1}\,  \alpha}{4\sqrt{1+\alpha^2}} \, .
\end{align}
It follows that the correlation achieves the bound:
\begin{align}
    &P(a_1 = b \mid x_1 = x_*, c = 0) + \alpha\, P(a_2 = x_1 \mid b = b_*, c = 1) \nonumber \\
    &= \frac 1 2 \sum_{a_1,a_2,x_1,b} \delta_{a_1,b}\, \delta_{x_1,x_*}\, p(a_1,a_2 \mid x_1, b,0) \nonumber \\
    &\quad + \frac {\alpha} 2 \sum_{a_1,a_2,x_1,b} \delta_{a_2,x_1}\, \delta_{b,b_*}\, p(a_1,a_2 \mid x_1, b,1) \nonumber \\
    &= \frac 1 2 \left( 1+ \frac{1}{\sqrt{1+\alpha^2}} \right) + \frac {\alpha} 2 \left( 1+ \frac{\alpha}{\sqrt{1+\alpha^2}} \right) \nonumber \\
    &= \frac{1}{2} \left( 1+\alpha + \sqrt{1+\alpha^2} \right) \, .
\end{align}
\qed

\section{Biased LGYNI correlations}
\label{app:tiltedlgyni}

% \begin{figure}[htbp]
%     \centering
%     \includegraphics{plot.pdf}
%     \caption{Causal bound, ICO bound and algebraic maximum of biased LGYNI (\ref{eq:tilted_lgyni}).  {\color{blue}     Tsirelson  $\to$  ICO,  Algebraic  $\to $ algebraic}}
%     \label{fig:plot}
% \end{figure}

\begin{figure*}[htbp]
    \centering
    \includegraphics[width=0.9\textwidth]{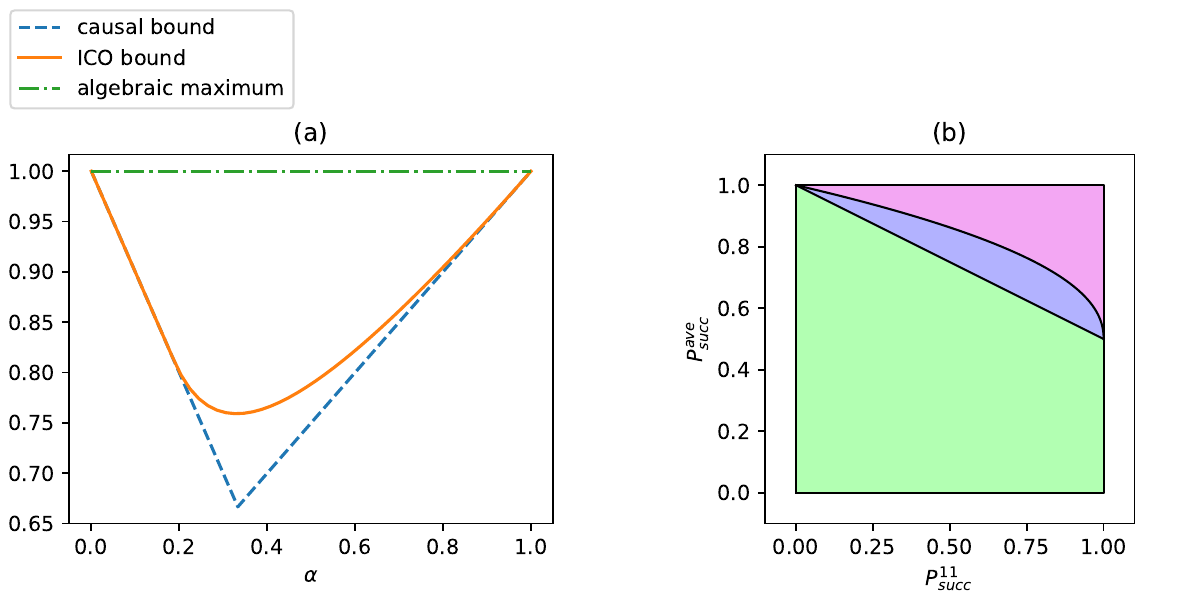}
    \caption{{\bf Visualizing correlations and probability distributions.} (a) Causal bound, ICO bound and algebraic maximum of the success probability of biased LGYNI. (b) Geometry of probability distributions on a 2-dimensional plane. The green area represents the projection of causal probability distributions; the blue area represents the projection of probability distributions exhibiting ICO. The remaining area (violet area) represents the non-causal probability distributions beyond the ICO ones.}
    \label{fig:lgynigeometry}
\end{figure*}

  We now introduce a biased version of the LGYNI game, corresponding to the correlation 
\begin{equation}
    \label{eq:tilted_lgyni}
\map I_{{\rm LGYNI}, \alpha}    =      \alpha\,  P_{\rm succ}^{11}   +  (1-\alpha)  \,  P_{\rm succ}^{\rm ave} \,,
\end{equation}
where $\alpha$ ranges between 0 and 1 and $P_{\rm succ}^{\rm ave} := \frac{  P_{\rm succ}^{01}   +   P_{\rm succ}^{10}    }2$ is the average of $P_{\rm succ}^{01}$ and $P_{\rm succ}^{10}$.     Clearly, the correlation $\map I_{{\rm LGYNI}, \alpha}$ is single-trigger, and for $\alpha= 1/3$ its maximization is equivalent to the maximization of the probability of success in the original LGYNI  game.     

For the biased LGYNI game, one has the causal inequality   
\begin{align}\label{LGYNIcausalineq}
\map I_{{\rm LGYNI}, \alpha}  \le       \max \left\{ 1-\alpha, \frac{1+\alpha}2 \right\}  \,. 
\end{align} 
The proof is as follows: let $p_A$ and $p_B$ be the marginal probability distributions of Alice's and Bob's outcomes, defined as
\begin{align}
    p_A(a_1 \mid x_1,x_2) &:= \sum_{a_2} p(a_1,a_2 \mid x_1,x_2) \, , \\
    p_B(a_2 \mid x_1,x_2) &:= \sum_{a_1} p(a_1,a_2 \mid x_1,x_2)
\end{align}
respectively.
If Alice's operation causally precedes Bob's, then
\begin{equation}
    \label{eq:causalboundlgyni_1}
    p(1,1 \mid 1,1) \leq p_A(1 \mid 1,1) = p_A(1 \mid 1,0)
\end{equation}
In this case, biased LGYNI is upper bounded as follows
\begin{align}
    &\map I_{{\rm LGYNI}, \alpha} \nonumber \\
    &= \alpha\,  P_{\rm succ}^{11}   +  (1-\alpha)  \,  \frac{  P_{\rm succ}^{01}   +   P_{\rm succ}^{10}    }2 \nonumber \\
    &=\alpha p(1,1 \mid 1,1) + \frac{1-\alpha} 2 p_A(0 \mid 1,0) + \frac{1-\alpha} 2 p_B(0 \mid 0,1) \nonumber \\
    &\leq \alpha p_A(1 \mid 1,0) + \frac{1-\alpha} 2 p_A(0 \mid 1,0) + \frac{1-\alpha} 2 p_B(0 \mid 0,1) \nonumber \\
    &= \alpha (1-p_A(0 \mid 1,0)) + \frac{1-\alpha} 2 p_A(0 \mid 1,0) \nonumber \\
    &\quad + \frac{1-\alpha} 2 p_B(0 \mid 0,1) \nonumber \\
    &\leq \max\left\{ \alpha, \frac{1-\alpha} 2 \right\} + \frac{1-\alpha} 2  \nonumber \\
    &= \max\left\{ \frac{1+\alpha} 2, 1-\alpha \right\}  \, , \label{eq:causalboundlgyni}
\end{align}
where the first inequality is due to (\ref{eq:causalboundlgyni_1}), and the second inequality is due to the monotonicity as a linear function of $p_A(0 \mid 1,0)$ and the probability constraint $p_B \leq 1$. 
Vice versa, biased LGYNI is upper bounded by the same value when Bob's operation precedes Alice's. Since a random mixture can not increase the bound, the causal bound of biased LGYNI is no greater than  (\ref{eq:causalboundlgyni}). 
The value (\ref{eq:causalboundlgyni}) can be saturated by the following causal process: Alice sends her classical bit $x$ to Bob; she produces outcome 0 if $\alpha < \frac{1-\alpha} 2$ and produces outcome 1 otherwise; Bob produces his outcome to be $x$.

Quantum processes with ICO can violate the inequality  (\ref{LGYNIcausalineq}) for suitable values of the    parameter $\alpha$.  Explicitly, the ICO bound for the biased LGYNI game can be computed as an SDP. The result   is shown in Figure \ref{fig:lgynigeometry}(a)  for all possible values of $\alpha$ in the interval $[0,1]$.      Interestingly, it appears that the biased LGYNI causal inequality cannot be violated by any quantum ICO process  for $\alpha \le  0.188$. (In the last part of this section, we provide a rigorous proof of this fact for $\alpha \leq \frac{4-\sqrt 5}{11} \approx 0.16$. ) 
  To the best of our knowledge, biased LGYNI with $0 < \alpha \le  0.188$   is the first non-trivial example of  a causal inequality with no quantum violation in the ICO framework.  

At first sight, the inequality  (\ref{LGYNIcausalineq}) for $\alpha \le  0.188$  
 may appear as a causal  analogue of Bell inequalities with no quantum violations \cite{almeida2010guess,fritz2013local}.    A more in-depth analysis of the geometry of ICO correlations, however, reveals  that this inequality is not tight, meaning that it does not identify a facet of the quantum ICO set.

 Let us  visualize the  success probabilities $P_{\rm succ}^{11} $ and $P_{\rm succ}^{\rm ave}$ in a two-dimensional plane.     The values of $P_{\rm succ}^{11} $ and $P_{\rm succ}^{\rm ave}$ compatible with the causal inequality (\ref{LGYNIcausalineq})  and with the ICO bound are shown in Figure \ref{fig:lgynigeometry}(b), where they correspond to the green area and the blue area, respectively.   The constraints arising from the causal inequality can be equivalently characterized  by the condition
\begin{equation}
    \label{eq:polytope2}
    P_{\rm succ}^{11} + 2P_{\rm succ}^{\rm ave} \leq 2 \, ,\\
\end{equation}
The condition (\ref{eq:polytope2}) is the causal inequality associated with the canonical LGYNI game, which corresponds to a facet of the causal polytope \cite{branciard2015simplest}.  
% See the proof of the characterizations (\ref{eq:polytope1}) and (\ref{eq:polytope2}) in Supplementary Note 5.

The shape of the boundary of the quantum ICO set, instead, shows that the set of probability distributions generated by quantum ICO processes  is not a polytope, {\em i.e.}  it has infinite extreme points.      Figure \ref{fig:lgynigeometry}(b) also shows that the boundary of the quantum ICO set is distinct from the boundary of the causal polytope for every value except for the extreme points of the causal polytope, corresponding to the values $p_{\rm succ}^{11}  =  0$ and  $p_{\rm succ}^{11}  =  1$.  The fact that the quantum ICO and causal boundaries do not coincide implies that the causal inequality (\ref{LGYNIcausalineq}) is not tight.    An interesting open question is whether there exist examples of tight causal inequalities with no quantum ICO violation, namely whether there exists flat faces of the quantum set that coincide with faces of the causal polytope.     

% In this section, we generalize the correlation of LGYNI game to a tilted version and investigate its causal bound and ICO bound. In addition, we show that when the parameter $\alpha$ is small enough, the ICO bound of biased LGYNI is equal to its causal bound.

Now we rigorously prove that when $\alpha \in \left[0, \frac{4-\sqrt 5}{11}\right]$, the ICO bound of biased LGYNI is equal to its causal bound $1-\alpha$. We claim that the follow positive operator $C$ on a 6-qubit Hilbert space $(\spc H_1 \otimes \spc H_{2,3}) \otimes (\spc H_4 \otimes \spc H_{5,6})$ is equal to $(1-\alpha)$ times the Choi operator of a  no-signalling channel
\begin{equation}
    C := \Omega + |\Psi_{01}\>\<\Psi_{01}| + |\Psi_{10}\>\<\Psi_{10}| + |\Psi_{00}\>\<\Psi_{00}| \, ,
\end{equation}
where $\Omega$ is the single-trigger operator of biased LGYNI
 \begin{align}
    \Omega &:=\, \alpha |11\>\< 11| \otimes |1\>\<1| \otimes |11\>\<11| \otimes |1\>\<1| \nonumber \\
    &\, + \frac{1-\alpha}2 \, |00\>\< 00| \otimes |1\>\<1| \otimes |I\kk\bb I| \otimes |0\>\<0| \nonumber \\
    &\, + \frac{1-\alpha}2 \, |I\kk\bb I| \otimes |0\>\<0| \otimes |00\>\<00| \otimes |1\>\<1| \, ,
\end{align}
and
\begin{align}
     |\Psi_{01}\> &:= c_0 |11\> \otimes |0\> \otimes |11\> \otimes |1\> \\
     & \quad + c_1 |00\> \otimes |0\> \otimes |11\> \otimes |1\> \nonumber \\
     |\Psi_{10}\> &:= c_0 |11\> \otimes |1\> \otimes |11\> \otimes |0\> \\
     & \quad + c_1 |11\> \otimes |1\> \otimes |00\> \otimes |0\> \nonumber \\
     |\Psi_{00}\> &:= c_2 |11\> \otimes |0\> \otimes |11\> \otimes |0\> \\
     & \quad + c_3 |00\> \otimes |0\> \otimes |11\> \otimes |0\> \nonumber \\
     & \quad + c_3 |11\> \otimes |0\> \otimes |00\> \otimes |0\> \nonumber \, ,
\end{align}
with non-negative coefficients $c_0, \cdots, c_3$. 
The conditions that $C$ is proportional to the Choi operator of a no-signalling channel are
\begin{equation}
    \Tr_{2,3} C = \frac{I_1} 2 \otimes \Tr_{1,2,3}C \, , \quad
    \Tr_{5,6} C = \frac{I_4} 2 \otimes \Tr_{4,5,6}C \, ,
\end{equation}
which is equivalent to the following equations:
\begin{align}
    c_1^2 - c_0^2 &= \alpha \, , \label{eq:coeff1} \\
    c_0^2 + c_2^2 - c_3^2 &= \frac{1-\alpha}2 \, , \\
    c_1^2 + c_3^2 &= \frac{1-\alpha}2 \, . \label{eq:coeff3} \\
    c_0c_1 + c_2c_3 &= \frac{1-\alpha}2 \, , \label{eq:coeff4}
\end{align}
We can solve $c_1$, $c_2$ and $c_3$ in terms of $c_0$ according to Eqs. (\ref{eq:coeff1})-(\ref{eq:coeff3})
\begin{align}
    c_1 &= \sqrt{c_0^2 + \alpha} \, , \label{eq:solvec1} \\
    c_2 &= \sqrt{1-2\alpha - 2c_0^2} \, , \\
    c_3 &= \sqrt{\frac{1 - 3\alpha}2 - c_0^2} \, . \label{eq:solvec3}
\end{align}
It follows that
\begin{equation}
    \frac{\Tr(C)} 4 = \frac{\alpha + 2(1-\alpha) + 2 \left(c_0^2 + c_1^2 \right) + \left(c_2^2 + 2c_3^2 \right)} 4 = 1-\alpha
\end{equation}
according to Eqs. (\ref{eq:solvec1})-(\ref{eq:solvec3}). Now it suffices to show the existence of $c_0, \cdots, c_3$ satisfying Eqs. (\ref{eq:coeff1})-(\ref{eq:coeff4}). With the solutions in Eqs. (\ref{eq:solvec1})-(\ref{eq:solvec3}), we focus on the quantity $f(c_0) = c_0c_1 + c_2c_3 - \frac{1-\alpha}2$, corresponding to Eq. (\ref{eq:coeff4}). When $\alpha \in \left[0, \frac{4-\sqrt 5}{11} \right]$, it holds that
\begin{equation}
    f(0) = \sqrt{\frac{(1-2\alpha)(1 - 3\alpha)}2} - \frac{1-\alpha}2 \geq 0 \, ;
\end{equation}
and
\begin{equation}
    f\left(\sqrt{\frac{1 - 3\alpha}2} \right) = \sqrt{\frac{1 - 3\alpha}2}\sqrt{\frac{1 - \alpha}2} - \frac{1-\alpha}2 \leq 0 \, ;
\end{equation}
By continuity of the function $f$, there exists $c_0 \in \left[0, \sqrt{\frac{1 - 3\alpha}2} \right]$ such that $f(c_0) = 0$ and thus Eq. (\ref{eq:coeff4}) is satisfied. Therefore, $(1-\alpha)$ is an upper bound of $\eta(\Omega)$ and thus an upper bound of the biased LGYNI. This completes the proof.

\section{Characterization of supermaps in bistochastic quantum and classical theories}
\label{app:classify}

\begin{figure*}
    \includegraphics{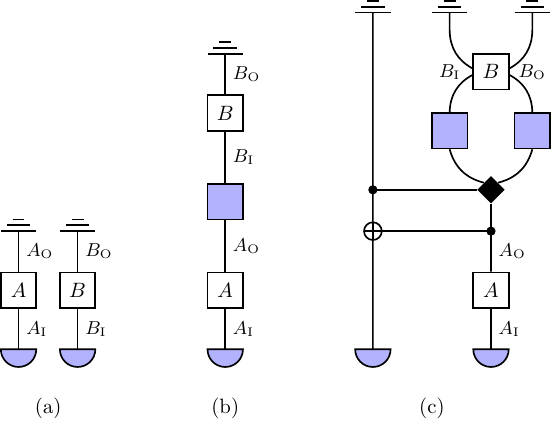}
    \caption{{\bf Circuit realization of simplest classical supermaps with definite causal order.} (a) no-signalling supermaps ($A \not\preceq B$ and $B \not\preceq A$),
    (b) unidirectional signalling supermaps ($A \preceq B$ or $B \preceq A$) with fixed time direction,
    (c) unidirectional signalling supermaps ($A \preceq B$ or $B \preceq A$) with dynamical time direction. In this figure, the blue half discs correspond to classical bits, and the blue boxes correspond to transformations that are either identity or bit-flip.
    }
    \label{fig:causal}
\end{figure*}

In this section, we show the proof of the characterization of quantum supermaps transforming a pair of bistochastic channels. We then provide a classification of the set of simplest bipartite classical supermaps, considering the classical limits of bistochastic quantum theory. This set includes the process used to achieve a guaranteed win in the GYNI game, as discussed in the main text.

\subsection*{Characterization of quantum supermaps}

We consider a supermap transforming a pair of bistochastic channels, one mapping states on system $A_{\rm I}$ to to states on system $A_{\rm O}$ and the other mapping states on system $B_{\rm I}$ to states on system $B_{\rm O}$. The output of the supermap is a channel from system $C_{\rm I}$ to system $C_{\rm O}$. Since a supermap is a completely positive map, its Choi operator $S$ has to be a positive operator. The requirement that $S$ transforms any pair of bistochastic channels into an ordinary channel is equivalent to the normalization condition
\begin{equation}
    \label{eq:supermapnormalize}
    \Tr[(\rho_{C_{\rm I}} \otimes I_{C_{\rm O}} \otimes M_{A_{\rm I}A_{\rm O}} \otimes N_{B_{\rm I}B_{\rm O}})^T S] = 1 \, ,
\end{equation}
where $\rho$ is an arbitrary density operator on system $C_{\rm I}$, and $M$ ($N$) is the Choi operator of an arbitrary bistochastic channel from $A_{\rm I}$ ($B_{\rm I}$) to $A_{\rm O}$ ($B_{\rm O}$). Specifically, as the Choi operator $M$ can be written as \cite{chiribella2022quantum}
\begin{equation}
    M = \frac{I_{A_{\rm I}} \otimes I_{A_{\rm O}}}{d_A} + T
\end{equation}
where $T$ is an operator such that $\Tr_{A_{\rm I}}T = 0$ and $\Tr_{A_{\rm O}}T = 0$,
the subspace spanned by $M$ is characterized by the projection $\Pi_A$
\begin{align}
    \Pi_A(M) &:= {}_{[A_{\rm I}A_{\rm O}]}M + {}_{[1-A_{\rm I}-A_{\rm O}+A_{\rm I}A_{\rm O}]}M \nonumber \\
    &= {}_{[1-A_{\rm I}-A_{\rm O}+2A_{\rm I}A_{\rm O}]}M \, ,
\end{align}
where we have used the notation $_{[X]}S := \Tr_X[S] \otimes \frac {I_X}{d_X}$ introduced in Methods.
Similarly, the subspace spanned by $N$ is characterized by the projection $\Pi_B$
\begin{equation}
    \Pi_B(N) := {}_{[1-B_{\rm I}-B_{\rm O}+2B_{\rm I}B_{\rm O}]}N \, .
\end{equation}
% Let $\spc L_0(X)$ denote the subspace spanned by the identity operator on system $X$, and let $\spc L_1(X)$ denote the subspace spanned by traceless operators on system $X$. 
% The set of all operators on system $X$ is denoted by $\spc L(X)$. 
% In condition (\ref{eq:supermapnormalize}), it turns out that the density operators $\rho$ span $\spc L(C_{\rm I})$, the operators $M$ span $\map L_0(A_{\rm I}) \otimes \map L_0(A_{\rm O}) \oplus \map L_1(A_{\rm I}) \otimes \map L_1(A_{\rm O})$,  and the operators $N$ span $\map L_0(B_{\rm I}) \otimes \map L_0(B_{\rm O}) \oplus \map L_1(B_{\rm I}) \otimes \map L_1(B_{\rm O})$. Let $\Pi$ be the projection onto the tensor product of these three subspace (spanned by $\rho$, $M$ and $N$ respectively), and 
Let $S'$ be the the reduced operator $S' := \Tr_{C_{\rm O}}S$. It follows that condition (\ref{eq:supermapnormalize}) is equivalent to
\begin{align}
    \Pi_A \otimes \Pi_B (S') = \frac{I_{A_{\rm I}A_{\rm O}B_{\rm I}B_{\rm O}C_{\rm I}}}{d_Ad_B} \, ,
\end{align}
which is in turn equivalent to the following two conditions
\begin{align}
    &\Tr(S) = d_Ad_Bd_{C_{\rm I}} \, , \label{eq:supermaptr1} \\
    &\Pi_A \otimes \Pi_B (S') = {}_{[A_{\rm I}A_{\rm O}B_{\rm I}B_{\rm O}C_{\rm I}]}S' \, . \label{eq:supermapspace}
\end{align}
Eq. (\ref{eq:supermapspace}) can be decomposed into four constraints
\begin{align}
    &{}_{[A_{\rm I}A_{\rm O}B_{\rm I}B_{\rm O}(1-C_{\rm I})C_{\rm O}]}S = 0 \, , \label{eq:supermapcons3}   \\
    &{}_{[(1-A_{\rm I})(1-A_{\rm O})B_{\rm I}B_{\rm O}C_{\rm O}]}S = 0 \, , \nonumber \\
    &{}_{[(1-B_{\rm I})(1-B_{\rm O})A_{\rm I}A_{\rm O}C_{\rm O}]}S = 0 \, , \nonumber \\
    &{}_{[(1-A_{\rm I})(1-A_{\rm O})(1-B_{\rm I})(1-B_{\rm O})C_{\rm O}]}S = 0 \, , \nonumber
\end{align}
which can be regarded as commuting projections. This completes the proof. The characterization presented in Methods is a special case of (\ref{eq:supermaptr1}) and (\ref{eq:supermapcons3}) with trivial systems $C_{\rm I}$ and $C_{\rm O}$.

\subsection*{Simplest bipartite classical supermaps}

In the simplest case, the systems $A_{\rm I}$, $A_{\rm O}$, $B_{\rm I}$ and $B_{\rm O}$ are all qubits, while the systems $C_{\rm I}$ and $C_{\rm O}$ are trivial. Classical supermaps can be represented by operators diagonal in the computational basis of $\spc H_{A_{\rm I}}\otimes \spc H_{A_{\rm O}} \otimes \spc H_{B_{\rm I}} \otimes \spc H_{B_{\rm O}}$
\begin{equation}
    \label{eq:simplestclassicalprocess}
    S = \sum_{i,j,k,l \in \{ 0,1 \}} S_{i,j,k,l} \, |i\>\<i| \otimes |j\>\<j| \otimes |k\>\<k| \otimes |l\>\<l| \, .
\end{equation}
Theorem 4 in Methods implies that the coefficients in Eq. (\ref{eq:simplestclassicalprocess}) are probabilities and satisfy
\begin{equation}
    \begin{split}
        \sum_{i=j, \, k=l} S_{i,j,k,l} &= 1 \, , \\
        \sum_{i=j, \, k=l\oplus 1} S_{i,j,k,l} &= 1 \, , \\
        \sum_{i=j\oplus 1, \, k=l} S_{i,j,k,l} &= 1 \, , \\
        \sum_{i=j\oplus 1, \, k=l\oplus 1} S_{i,j,k,l} &= 1 \, .
    \end{split}
\end{equation}
It follows that these classical supermaps form a convex set with $2^8$ extreme points:
\begin{equation}
    \label{eq:supermaps}
    \sum_{m=0}^3 |i_m\>\<i_m| \otimes |j_m\>\<j_m| \otimes |k_m\>\<k_m| \otimes |l_m\>\<l_m|
\end{equation}
where
\begin{align}
    &i_0=j_0, \, k_0=l_0 \, , \\
    &i_1=j_1, \, k_1=l_1\oplus 1 \, , \\
    &i_2=j_2\oplus 1, \, k_2=l_2 \, , \\
    &i_3=j_3\oplus 1, \, k_3=l_3\oplus 1 \, .
\end{align}

We provide a classification of the $2^8$ extreme points, corresponding to deterministic classical supermaps without a pre-defined direciton of time:
\begin{itemize}
    \item No-signalling supermaps ($A \not\preceq B$ and $B \not\preceq A$). There are $2^4$ points of this type, all equivalent to the circuit in Figure \ref{fig:causal}(a), up to interchange of input/output systems;
    \item Unidirectional signalling supermaps ($A \preceq B$ or $B \preceq A$) with fixed time direction. There are $2^5$ points of this type, all equivalent to the circuit in Figure \ref{fig:causal}(b), up to interchange of input/output systems and causal orders;
    \item Unidirectional signalling supermaps ($A \preceq B$ or $B \preceq A$) with dynamically controlled time direction. There are $2^6$ points of this type, all equivalent to the circuit in Figure \ref{fig:causal}(c) where the time direction of $B$ is dynamically decided by the output of $A$, up to interchange of input/output systems and causal orders;
    \item Supermaps with indefinite causal order and time direction. There are $2^4 + 2^7$ points of this type. Up to interchange of input/output systems, $2^4$ points of them can be expressed through functions: $i_m = j_m \oplus l_m \oplus \alpha$ and $k_m = j_m \oplus l_m \oplus \beta$ where $\alpha,\beta \in \{0,1\}$. The other $2^7$ points, however, cannot be understood as stochastic processes.
\end{itemize}

The classical supermap used to achieve a guaranteed win in the GYNI
game in the main text corresponds to the extreme point given by:
\begin{equation}
    \label{eq:coeffbisig}
    S_{0,0,0,0} = S_{1,1,1,0} = S_{1,0,1,1} = S_{0,1,0,1} = 1 \, .
\end{equation}

\section{Reaching the algebraic maximum in certain scenarios}
\label{app:omnipotent}

In this section, we provide constructions of strategies in bistochastic classical theory which realize algebraically maximal correlations. Let $(N,n,m)$ denote the scenario with $N$ parties, each with $n$ settings and $m$ outcomes. We show that perfect two-way signalling of ternary digits can be achieved with choices of classical supermaps and local bistochastic instruments. Equipped with a perfect two-way signalling of each other's setting, the two parties can generate arbitrary deterministic conditional probability distribution and thus the BICO value in the (2,3,?) scenario is equal to the algebraic maximum. Moreover, we show that if there are only 2 settings for each party (the (2,2,?) scenario), every bipartite probability distribution can be generated with choices of classical supermaps and local bistochastic instruments. Our constructions cannot be directly extended to the scenarios with more than 3 settings. This leads us to raise the open question whether there are non-trivial constraints on the correlations generated in bistochastic classical and quantum theory in general scenarios.

Let us consider bipartite classical supermaps transforming a pair of classical bistochastic instruments to probabilities. These supermaps can be represented by operators $S$ diagonal in an orthonormal basis on the Hilbert space $\spc H_{A_{\rm I}}\otimes \spc H_{A_{\rm O}} \otimes \spc H_{B_{\rm I}} \otimes \spc H_{B_{\rm O}}$ corresponding to Alice's and Bob's input/output systems
\begin{equation}
    \label{eq:classicalsupermap}
    S = \sum_{i,j,k,l \in \{ 0,...,d-1 \}} S_{i,j,k,l} \, |i\>\<i| \otimes |j\>\<j| \otimes |k\>\<k| \otimes |l\>\<l| \, 
\end{equation}
where $d$ is the dimension of every input/output system and $S_{i,j,k,l}$ are probabilities.

\subsection*{Perfect two-way signalling in the (2,3,?) scenerio}

In the scenario where each of the two parties has $n\leq 3$ settings, there is a strategy generalizing the one introduced in the main text for winning the GYNI game, which realizes two-way signalling of $n$-nary digits. In this strategy, Alice uses the classical instrument
\begin{equation*}
    q^{(1)}_{a_1 \mid x_1}(s'_1 \mid s_1) = \delta_{a_1,s'_1}\delta_{s_1', ((s_1 - x_1)\mod n)} \, ,
\end{equation*}
and Bob uses the classical instrument
\begin{equation*}
    q^{(2)}_{a_2 \mid x_2}(s'_2 \mid s_2) = \delta_{a_2,s'_2}\delta_{s_2', ((x_2 - s_2)\mod n)} \, .
\end{equation*}
Alice's and Bob's local experiments are connected by a
deterministic classical supermap, represented as a function that maps the final values of the classical systems $(s'_1, s'_2)$ into their initial values $(s_1, s_2)$ according to the rule
\begin{equation}
    \label{eq:supermapn}
    s_1 = ((s'_1 + s'_2)\mod n) \, , \quad s_2 = ((s'_1 - s'_2)\mod n) \, .
\end{equation}
The classical supermap $(\ref{eq:supermapn})$ can be represented in the form of Eq. (\ref{eq:classicalsupermap}), with non-zero coefficients given by $S_{i,j,k,l} = 1$ for $j,l \in \{0,1,\cdots, n-1\}$, $i=((j+l)\mod n)$ and $k=((j-l)\mod n)$. We can check that it satisfies the conditions (\ref{eq:supermaptr1}) and (\ref{eq:supermapcons3}) when $n\leq 3$. However, when $n = 4$, the map (\ref{eq:supermapn}) is not an admissible supermap. 

\subsection*{The (2,2,?) scenario}

If each party has 2 settings, we can construct a class of classical supermaps which not only enable perfect signalling of bits but also distribute shared randomness between the two parties. This allows the two parties to create any random mixture of deterministic conditional probability distributions. The class of supermaps, denoted by $\{\map S_{x,y}\}$, can be represented by positive operators $\{ S_{x,y} \}$ on the Hilbert space $\spc H_{A_{\rm O}}\otimes \spc H_{B_{\rm O}} \otimes \spc H_{A_{\rm I}} \otimes \spc H_{B_{\rm I}}$:
\begin{align}
    S_{x,y} &:= (|x,x\>\<x,x| + |y,y\>\<y,y|)\otimes (I-|y\>\<y|) \otimes |x\>\<x| \nonumber \\
    & + (|x,y\>\<x,y| + |y,x\>\<y,x|) \otimes |y\>\<y| \otimes (I-|x\>\<x|) \nonumber \\
    & + (I-|x\>\<x|-|y\>\<y|)^{\otimes 2} \otimes |y\>\<y| \otimes |x\>\<x| \label{eq:sxy}
\end{align}
where $x,y\in \{0,\cdots, d-1\}$. We can check that the positive operators (\ref{eq:sxy}) satisfy the conditions (\ref{eq:supermaptr1}) and (\ref{eq:supermapcons3}). Assuming that $d$ is even, we construct the classical bistochastic instrument of party $i$ to be a deterministic function $q^{(i)}_{a_i\mid x_i}(s'_i \mid s_i)$ which gives the outcome $a_i = f_i^{(t)}(x_i, z)$ and produces the classical state $s'_i = 2t + z$, with $t = \lfloor \frac{s_i}2 \rfloor$, $z=((x_i + s_i)\mod 2)$ and $f_i^{(t)}$ is an function depending on the parameter $t \in \{0, \cdots, \frac d 2-1\}$. If the local experiments are connected by the supermap $\map S_{2t, 2t+1}$, then the correlation between the two parties is the following deterministic probability distribution
\begin{equation}
    p(a_1,a_2 \mid x_1,x_2) = \delta_{a_1, f_1^{(t)}(x_1, x_2)} \delta_{a_2, f_2^{(t)}(x_2, x_1)} \, .
\end{equation}
An arbitrary random mixture of deterministic probability distributions can be generated by randomly choosing the supermap from $\{\map S_{2t,2t+1}\}$. In other words, the above strategy can be used to generate arbitrary bipartite probability distribution in the (2,2,?) scenario.

\section{Proof of Theorem 1 in Methods}
\label{app:labelledproj}

Consider the realization of the instruments $\map M_{a_i \mid x_i}^{(i)}$ by an isometry and a projective measurement.  Let $\left\{ K_{a_i\mid x_i,j}^{(i)} \right\}$ be a Kraus decomposition of the instrument $\map M_{a_i \mid x_i}^{(i)}$ of the $i$-th party. For every classical label $x_i$, define $V_{x_i}^{(i)}$ to be the following isometry from $\spc H_{A_{\rm I}^{(i)}}$ to $\spc H_L \otimes \spc H_{A_{\rm O}^{(i)}} \otimes \spc H_J$
\begin{equation}
    V^{(i)}_{x_i} := \sum_{a_i, j} |a_i\>_L \otimes K_{a_i\mid x_i,j}^{(i)} \otimes |j\>_J \, ,
\end{equation}
where $L$ is the classical register for the outcome $a_i$ and $J$ the classical register for the index $j$. Choosing the associated Hilbert spaces to be $\spc H_L \simeq \C^{m_i}$ and $\spc H_{J} \simeq \spc H_{A_{\rm I}^{(i)}} \otimes \spc H_{A_{\rm O}^{(i)}}$, the two systems $L$ and $J$ have no dependence on the setting $x_i$.  Extending $V_{x_i}^{(i)}$ to an unitary $U_{x_i}^{(i)}$ from $\spc H_{A_{\rm I}^{(i)}} \otimes \spc H_{E_{\rm I}^{(i)}}$ to $\spc H_L \otimes \spc H_{A_{\rm O}^{(i)}} \otimes \spc H_J \otimes \spc H_{E_{\rm O}^{(i)}}$ with some auxiliary systems $E_{\rm I}^{(i)}$ and $E_{\rm O}^{(i)}$, the map $\map M_{a_i \mid x_i}^{(i)}$ can be realized as
\begin{equation}
    \label{eq:unitaryrealization}
    \rho \mapsto \Tr_{L,J,E_{\rm O}^{(i)}} \left[\Pi_{a_i} U_{x_i}^{(i)} \left(\rho \otimes |0\>\<0|_{E_{\rm I}^{(i)}} \right) {U_{x_i}^{(i)}}^\dagger \Pi_{a_i} \right] \, ,
\end{equation}
where
\begin{equation}
    \Pi_{a_i} = |a_i\>\<a_i|_L \otimes I_{A_{\rm O}^{(i)}} \otimes I_{J} \otimes I_{E_{\rm O}^{(i)}} \, .
\end{equation}
Define local operations $\map E^{(i)}$ and $\map D^{(i)}$ to be the channels
\begin{align}
    \forall \rho \quad \map E^{(i)}(\rho) &= \rho \otimes |0\>\<0|_{E_{\rm I}^{(i)}} \, , \\
    \forall \sigma \quad \map D^{(i)}(\sigma) &= \Tr_{L,J, E_{\rm O}^{(i)}, F} \map U^{(i)}(\sigma) \, ,
\end{align}
where $\map U^{(i)}$ is the controlled unitary channel
\begin{equation}
    \sum_{x_i} U_{x_i}^{(i)} \otimes |x_i\>\< x_i|_{F} \, .
\end{equation}
with the control system denoted by $F$.

We construct the projectors of the labelled projective instruments $\map M_{a_i \mid x_i}^{(i)'}$ to be $\left\{ {U_{x_i}^{(i)}}^\dagger \Pi_{a_i} U_{x_i}^{(i)} \right\}_{a_i}$ (whose rank is independent of $a_i$ and $x_i$), the label of $\map M_{a_i \mid x_i}^{(i)'}$ to be $x_i$, and the process $\map S'$ to be the composite of the local operations $\map E^{(i)}$, $\map D^{(i)}$ and the process $\map S$, i.e.
\begin{align}
    &\map S'   \left(  \map M^{(1)'}_{a_1 \mid x_1}, \cdots, \map M^{(N)'}_{a_N \mid x_N}    \right) \label{eq:constructprocess} \\
    &= \map S \left(  \map D^{(1)} \circ \map M^{(1)'}_{a_1 \mid x_1} \circ \map E^{(1)}, \cdots, \map D^{(N)} \circ \map M^{(N)'}_{a_N \mid x_N} \circ \map E^{(N)}    \right) \nonumber
\end{align}
According to the realization in Eq. (\ref{eq:unitaryrealization}), the local operations $\map E^{(i)}$ and $\map D^{(i)}$ transform the labelled projective instruments $\map M_{a_i \mid x_i}^{(i)'}$ to $\map M_{a_i \mid x_i}^{(i)}$. Hence $\map S'   \left(  \map M^{(1)'}_{a_1 \mid x_1}, \cdots, \map M^{(N)'}_{a_N \mid x_N}    \right)$ reproduces the joint conditional probability $\map S   \left(  \map M^{(1)}_{a_1 \mid x_1}, \cdots, \map M^{(N)}_{a_N \mid x_N}    \right)$.

\section{Proof of Theorem 2 in Methods}
\label{app:performanceop}

The key of the proof is the fact that a projective measurement followed by the measurement outcome being discarded can be written as a uniform mixture of unitary channels. Specifically, let $\{P_k\}_{k=1}^{n}$ be the operators of a projective measurement. Proposition 4.6 of Watrous' textbook \cite{watrous2018theory} implies that there exist $2^{n}$ unitary gates $\{U_l\}_{l=1}^{2^n}$ such that
\begin{equation}
    \label{eq:proj2unitary}
    \forall \rho\quad \sum_{k=1}^{n} P_k\rho P_k = \frac{1}{2^n} \sum_{l=1}^{2^n} U_l \rho U_l^\dagger \, .
\end{equation}

Consider the map $\map M_{\map I}$ associated to a given single trigger correlation $\sum_{\vec a,\vec x}\alpha_{\vec a,\vec x}\, p(\vec a \mid \vec x)$, with triggers $\vec\xi$ and the labelled projective
instruments $\map M_{a_i \mid x_i}^{(i)}(\cdot) = P_{a_i,x_i}^{(i)} \cdot P_{a_i,x_i}^{(i)} \otimes |x_i\>\<x_i|$. We observe that the value of $\map M_{\map I}$ is unchanged when substituting $\map M_{a_i \mid x_i}^{(i)}$ with $\frac  1 {m_i} \sum_{k=0}^{m_i-1} M_{k\mid x_i}^{(i)}$ for any party $i$ and any setting $x_i \neq \xi_i$. For example, let $x_1$  be a non-trigger setting of the first party, i.e. $x_1 \neq \xi_1$. Indeed, since $\alpha_{\vec a, \vec x}$ does not depends on $a_1$, we have
\begin{align}
    \sum_{a_1=0}^{m_1-1} \alpha_{\vec a, \vec x} \map M_{a_1 \mid x_1}^{(1)}
    &= \frac 1 {m_1} \sum_{a_1=0}^{m_1-1} \sum_{k=0}^{m_1-1}  \alpha_{\vec a, \vec x}  \map M_{k \mid x_1}^{(1)} \nonumber \\
    &= \sum_{a_1=0}^{m_1-1} \alpha_{\vec a,\vec x} \left( \frac{\sum_{k=0}^{m_1-1} \map M_{k \mid x_1}^{(1)}} {m_1} \right) \, . 
\end{align}
According to the property in Eq. (\ref{eq:proj2unitary}), for every party $i$, there exists a collection of unitary channels $\left\{\map U_{x_i}^{(i,l)} \right\}_{l=1}^{2^{m_i}}$ such that
\begin{equation}
    \sum_{k=0}^{m_i-1} \map M_{k\mid x_i}^{(i)} = \frac 1 {2^{m_i}} \sum_{l=1}^{2^{m_i}} \map U_{x_i}^{(i,l)} \otimes |x_i\>\< x_i| \, .
\end{equation}
Iterating the above substitutions for every party $i$ and every setting $x_i \neq \xi_i$, $\map M_{\map I}$ is readily in the form of a uniform mixture of the maps $( \map M_{\map I,j} )_j$ presented in Theorem 2. This completes the proof.

\medskip
\section{Proof of ICO bound of single-trigger correlations}

% In fact, the elements of $\{ \Omega_j \}$ in Lemma \ref{theo:performanceop} in the main text  are equivalent to each other, up to no-signalling preserving unitary transformations. Explicitly, $\Omega_j = (I\otimes U_{x_i}^{(i,j)} \otimes I) \Omega_0 (I\otimes {U_{x_i}^{(i,j)}}^\dagger \otimes I)$, where $\Omega_0$ is the following operator

\label{app:isomorphism}

Theorem 1 and 2 in Methods imply that the optimization of a single-trigger correlation can be restricted without loss of generality to an optimization over the following instruments
\begin{equation}
    \map    M_{\,a_i|  x_i}^{(i)}  (\rho) = \begin{cases}
               P_{a_i,\xi_i}^{(i)} \, \rho  \,   P_{a_i,\xi_i}^{(i)} \otimes |\xi_i\>\< \xi_i| & x_i = \xi_i \, , \\
            \frac 1 {m_i}  \,  U_{x_i}^{(i)}  \,\rho \,  U_{x_i}^{(i)\dag}    \otimes |x_i\>\< x_i| & x_i \neq \xi_i \, .
        \end{cases}
    \end{equation}
where for each party $i$, $\left(P_{a_i,\xi_i}^{(i)} \right)_{a_i = 0}^{m_i-1}$ are projectors of equal rank and $\left(U_{x_i}^{(i)} \right)_{x_i \neq \xi_i}$ are unitary operators.  
The unitary operators can be further fixed without loss of generality to be the identity. Explicitly, $\map S$ being an arbitrary quantum process, the probability distribution $\map S   \left(  \map M^{(1)}_{a_1 \mid x_1}, \cdots, \map M^{(N)}_{a_N \mid x_N}    \right)$ can be reproduced by
\begin{align}
    &\map S'   \left( \map M^{(1)'}_{a_1 \mid x_1}, \cdots, \map M^{(N)'}_{a_N \mid x_N}    \right) \nonumber \\
    &= \map S   \left( \map U^{(1)} \circ \map M^{(1)'}_{a_1 \mid x_1}, \cdots, \map U^{(N)} \circ \map M^{(N)'}_{a_N \mid x_N}    \right) \, , \label{eq:applycontrolu}
\end{align}
where $\map M^{(i)'}_{a_i \mid x_i}$ are the instruments obtained from $\map M^{(i)}_{a_i \mid x_i}$ by letting $U_{x_i}^{(i)} = I$, $\map U^{(i)}$ is the controlled unitary gate $I \otimes |\xi_i\>\<\xi_i| + \sum_{x_i\neq \xi_i} U_{x_i}^{(i)} \otimes |x_i\>\<x_i|$, and the process $\map S'$ is the composite of the process $\map S$ and the local operations $\left(\map U^{(i)} \right)_{i=1}^N$.

The equal-rank projectors $\left(P_{a_i,\xi_i}^{(i)} \right)_{a_i = 0}^{m_i-1}$ can be written as the tensor product of rank-one projectors and the identity on an additional system $R^{(i)}$, i.e.
\begin{equation}
    P_{a_i,\xi_i}^{(i)} = |a_i\>\<a_i| \otimes I_{R^{(i)}} \, . \label{eq:torankone}
\end{equation}

Putting Eqs. (\ref{eq:applycontrolu}) and (\ref{eq:torankone}) together, the probability distribution $\map S   \left(  \map M^{(1)}_{a_1 \mid x_1}, \cdots, \map M^{(N)}_{a_N \mid x_N}    \right)$ can be written as
\begin{align}
    &\map S''   \left( \map M^{(1)*}_{a_1 \mid x_1}, \cdots, \map M^{(N)*}_{a_N \mid x_N}    \right) \nonumber \\
    &=\map S'   \left( \map M^{(1)*}_{a_1 \mid x_1} \otimes \map I_{R^{(1)}}, \cdots, \map M^{(N)*}_{a_N \mid x_N} \otimes \map I_{R^{(N)}}    \right) \, ,
\end{align}
where $\map M_{\,a_i|  x_i}^{(i)*}$ are the canonical instruments, and the process $\map S''$ is obtained by incorporating identity gates on the additional systems $R^{(i)}$ into the process $S'$. In other words, the maximum quantum ICO value of a single-trigger correlation can be obtained by fixing the local operation of each party to be the canonical instrument $\map M_{\,a_i|  x_i}^{(i)*}$.
This completes the proof.

\end{document}